\documentclass{article}

\usepackage[breaklinks=true,colorlinks=true,linkcolor=blue,urlcolor=blue,citecolor=blue]{hyperref}

\usepackage[utf8]{inputenc}
\usepackage[TS1,T1]{fontenc}
\usepackage{lmodern,microtype}
\usepackage{natbib}
\usepackage{epsfig,subfig}
\usepackage{amssymb,amsmath}
\usepackage{xcolor}
\usepackage{bm}
\usepackage[normalem]{ulem}

\usepackage{authblk}

\usepackage{lineno}
\modulolinenumbers[5]

\usepackage{geometry}
\newcommand{\aj}[1]{\textcolor{black}{#1}}

\newcommand{\ds}{\displaystyle}
\newcommand{\uu}[1]{\mathbf{#1}}
\newcommand{\tou}[1]{{\boldsymbol{#1}}}
\newcommand{\tod}[1]{{\boldsymbol{#1}}}
\newcommand{\moy}[2][]{{\left\langle{#2}\right\rangle}_{#1}}
\newcommand{\dd}{\,\text{d}}
\newcommand{\macr}[1]{{\overline{#1}}}
\newcommand{\abr}[1]{\text{\textsf{#1}}}
\newcommand{\eps}{\varepsilon}
\newcommand{\sig}{\sigma}

\graphicspath{{./figures/}}
\DeclareGraphicsExtensions{.pdf}

\title{{Numerical simulation of model problems in plasticity based on field dislocation mechanics}}

\author[,1,2]{Léo Morin\thanks{Corresponding author: \texttt{leo.morin@ensam.eu}}}
\author[3]{Renald Brenner}
\author[2]{Pierre Suquet}

\affil[1]{Laboratoire de M\'ecanique et d'Acoustique, {Aix-Marseille Univ, CNRS UMR 7031,} Centrale Marseille, 4 impasse Nikola Tesla, CS 40006, 13453 Marseille Cedex 13, France}
\affil[2]{Laboratoire PIMM, Arts et M\'etiers, CNRS, Cnam, HESAM Universit\'e, 151 Boulevard de l'H\^opital, 75013 Paris, France}
\affil[3]{Sorbonne Université, CNRS, UMR 7190, Institut Jean Le Rond $\partial$'Alembert, 75005 Paris, France}

\date{}

\begin{document}
\maketitle

\begin{abstract}
The aim of this paper is to investigate the numerical implementation of the Field Dislocation Mechanics (FDM) theory for the simulation of dislocation-mediated plasticity. First, \aj{the mesoscale} FDM theory \aj{of \cite{acharya-and-roy-2006}} is \aj{recalled} which permits to express the set of equations under the form of a static problem, corresponding to the determination of the local stress field for a given dislocation density distribution, \aj{complemented} by an evolution problem, corresponding to the transport of the dislocation density. The static problem is  solved using FFT-based techniques \citep{brenner_numerical_2014}. \aj{The main contribution of the present study is} an efficient numerical scheme based on high resolution Godunov-type solvers to solve the evolution problem. Model problems of dislocation-mediated plasticity are finally considered in a \aj{simplified layer case}. First, uncoupled problems with uniform velocity are considered, which permits to reproduce annihilation of dislocations and expansion of dislocation loops. \aj{Then, the FDM theory is applied to several problems of dislocation microstructures subjected to a mechanical loading.}\\
\end{abstract}

\vspace{2pc}


\section{Introduction}\label{sec:Intro}
{{Yielding and plastic deformation in crystalline materials at the single crystal scale is determined by 
underlying mechanisms at a smaller scale attached to the presence and to the motion of dislocations (line defects). }
The treatment of dislocations as discrete objects with local interacting rules (annihilation, junction formation, etc.) has led to the Discrete Dislocation
Dynamics (DDD) which dates back to the late eighties  {\citep{lepinoux_kubin_87,kubin_canova_92,giessen_1995}} (see \citet{kubin_13,po_ghoniem_14} for comprehensive reviews). The computational time-consuming part of this approach is the evaluation of the elastic interactions between all dislocation segments. }
\\
{
{A first way around this problem is the}  hybrid ``discrete-continuum'' approach \citep{lemarchand_2001} which makes use of the eigenstrain theory \citep{mura_1982}. In short, this model consists in an elastoplastic finite-element (FE) computation where the plastic flow rule is replaced by a DDD simulation. \\
{A different fully ``continuum'' approach,  consists in considering \aj{dislocation density field rather than individual dislocation segments}. Several  dislocation-mediated elastoplastic theories}, relating the elastic theory of continuously distributed dislocations \citep{willis_1967} to constitutive mesoscale plasticity models, have been proposed \aj{\citep{acharya_model_2001,acharya_driving_2003,acharya_constitutive_2004,roy_finite_2005,gurtin_2006, hochrainer_2014,xia_elazab_2015}.} {We follow here the  ``Field Dislocation Mechanics'' (FDM) of  \citet{acharya_model_2001} which is} a fully continuum model using the Nye dislocation tensor \citep{nye_1953} as an internal state variable field. \aj{At each point, it is related, in general, to dislocation lines bundles on different slip systems.} This {dislocation density}, linked to the incompatible part of the plastic distortion, allows {for the determination of} the internal stress state (i.e long-range elastic interactions) and the plastic strain rate can be derived from its evolution (transport equation) which expresses the conservation of the Burgers vector in the material.  \\
\aj{Our study is a contribution towards the derivation of a plasticity model able to describe size effects and dislocation patterning \citep{acharya-and-roy-2006,acharya-and-arora-2019}. More specifically, it is focused on the description of the plastic strain rate arising from the evolution of the dislocation density tensor field. This work builds upon a numerical study solely devoted to the numerical resolution of the internal stress field problem (i.e static FDM theory) for periodic media \citep{brenner_numerical_2014}.} \aj{This previous investigation has resorted to the numerical FFT scheme originally proposed by \citet{moulinec_numerical_1998} and now widely used for
micromechanical studies on the linear and nonlinear behaviors of heterogeneous materials. The uniqueness of the solution stress
field has been proven and an efficient numerical computational procedure for three-dimensional heterogeneous material with arbitrary elastic anisotropy has been proposed.} Interestingly, it can be noted that \citet{bertin_2015} {took advantage of}  this numerical approach to propose a dislocation dynamics model in line with \citet{lemarchand_2001}. It is worth mentioning the study of \cite{djaka_field_2017} which reports calculations of internal stress field, by using a FFT scheme, for various microstructural situations (see also \cite{berbenni_numerical_2014} for the homogeneous case). \\
	In the present article, we first use a rewriting of the transport equation in terms of the plastic distortion (Section \ref{sec:FDM}). \aj{A general procedure is then proposed to solve FDM plasticity
problems for which the plastic strain rate is only due to the evolution of the Nye tensor field (i.e. there is no contribution of statistically stored dislocations). It is possible to have recourse to phenomenological constitutive laws from classical crystal plasticity to handle this contribution \citep{acharya-and-roy-2006}. Note also that attempts have been proposed to derive
it from an average procedure of the behaviour of dislocations ensembles for 2D systems of straight edge dislocations \citep{groma_2003,valdenaire_2016}.} \aj{To neglect this contribution amounts to solve the evolution problem for the Nye tensor without source term.} \aj{This assumption is made in the present work and the transport equation is solved in Section \ref{sec:numeric} by means of a Godunov-type high resolution scheme which extends to dimension 2 the scheme of \cite{das_microstructure_2016}}. \\
Illustrative results are presented for simple problems with a constant dislocation velocity (Section \ref{sec:results_uncoupled}), namely annihilation and dislocation loop expansion which have been considered in previous studies, and finally model problems with a dependence of the velocity on the stress field are considered (Section \ref{sec:results_coupled}).
}

\section{Field Dislocation Mechanics theory}\label{sec:FDM}

\subsection{Primitive form of FDM}
The problem we are addressing is the numerical modelling of dislocation-mediated plasticity. The approach followed relies on the use of the Nye dislocation tensor field as internal state variable \citep{acharya_model_2001,acharya_constitutive_2004}. This requires to solve (i) a static problem, consisting in the determination of the internal stress field resulting from a given dislocation density field and an applied macroscopic stress in heterogeneous anisotropic elastic media and (ii) an evolution problem, consisting in the transport of the dislocation density field due to the local stress field produced.

The present study is focused on the case of an infinite medium with a periodic microstructure, that is, the Nye dislocation tensor $\bm{\alpha}$ and the elastic  moduli tensor $\uu{C}$ are considered as periodic fields. The problem {thus} consists {in} finding, for {a} given periodic dislocation field  $\bm{\alpha}$ and a macroscopic stress $\macr{\bm{\sigma}}$, the elastic distortion $\uu{U}^{\rm e}$, the stress $\bm{\sigma}$ and the rate of dislocation density $\dot{\bm{\alpha}}$ which solve, {on the unit-cell $V$},
\begin{align}\label{eq:EqFDM}
\left\{\begin{array}{lllr}
\uu{div}(\bm{\sigma}) & = & \uu{0} & {\rm Equilibrium~(static~problem)} \\
\bm{\sigma} & = & \uu{C}:\uu{U}^{\rm e} & {\rm Elasticity~law~ (static~problem)} \\
\uu{curl}(\uu{U}^{\rm e}) & = & \bm{\alpha} & {\rm Definition~of~ Nye~tensor~(static~problem)}  \\
\dot{\bm{\alpha}} & = & - \uu{curl}(\bm{\alpha} \times \uu{V}) & {\rm \quad \quad Transport~of~dislocation~(evolution~problem)},
\end{array}\right.
\end{align}
where $\uu{V}$ is the {dislocation} velocity whose constitutive relation needs to be specified. {The transport equation is the pointwise statement of the conservation law of the Burgers vector in the absence of source term.} The problem is {closed} by {periodic} boundary conditions {together with appropriate average relations}. {The use of periodicity conditions permits to avoid the extra-complication of boundary effects (such as free boundaries).}
From the definition of the Nye tensor \eqref{eq:EqFDM}$_3$, {it follows that}
\begin{equation}\label{eq:divalpha}
\uu{div}(\bm{\alpha}) = \uu{0}
\end{equation}
whose physical meaning is that dislocations cannot end within the material (they either form loops or reach the surface).

With in mind the numerical implementation and application of this constitutive model, {it is worth noting the following points:}
\begin{itemize}
\item \aj{The main kinematic variables of the model are the elastic distortion $\uu{U}^{\rm e}$ and the dislocation density $\bm{\alpha}$. The plastic part of the velocity gradient appears as $\bm{\alpha} \times \uu{V}$. The system of equation \eqref{eq:EqFDM} then allows for the determination of the displacement field.}
\item The last equation in \eqref{eq:EqFDM} is a transport equation for the dislocation density. As is well known in other problems involving conservation laws, \aj{the  numerical discretization of such systems} can only guarantee that the divergence condition is of order  of the numerical truncation error. In particular, the discrete divergence may become very large across shock waves and can lead to spurious solutions with unphysical oscillations (see \cite{rossmanith_unstaggered_2006}{, and references {herein},} in the context of magnetohydrodynamics flows). {A way to circumvent this issue is} to introduce a new variable accounting implicitly for the divergence condition.
\end{itemize}

\subsection{An elastoplastic formulation of FDM}
Based on the above remarks, we {adopt in the sequel} a slightly {reformulated} version of FDM
following \citet{acharya_new_2010}.

First, in order to introduce standard state variables, we recall  the multiplicative decomposition of the deformation gradient  $\uu{F}$:
\begin{equation}
\uu{F}=\uu{F}^{\rm e}.\uu{F}^{\rm p}
\end{equation}
where $\uu{F}^{\rm e}$ and $\uu{F}^{\rm p}$ are respectively the elastic and plastic part of the deformation gradient related to the elastic and plastic distortion $\uu{U}^{\rm e} $ and $\uu{U}^{\rm p}$:
\begin{equation}
\uu{F}^{\rm e} = \uu{I} + \uu{U}^{\rm e};\quad \quad \uu{F}^{\rm p} = \uu{I} + \uu{U}^{\rm p}.
\end{equation}
The deformation gradient being defined from the displacement field $\bm{u}$ via the relation $\uu{F} = \uu{I} + \nabla\bm{u}$, one can easily obtain
\begin{equation}
\nabla\bm{u} = \uu{U}^{\rm e} + \uu{U}^{\rm p} + \uu{U}^{\rm e}.\uu{U}^{\rm p}.
\end{equation}
This reduces, in small strains, to the following relation
\begin{equation}\label{eq:defUp}
\nabla\bm{u} = \uu{U}^{\rm e} + \uu{U}^{\rm p}
\end{equation}
which permits to express the elastic distortion as a function of the gradient of the displacement. Of course, the total strain is given by
\begin{equation}\label{eq:defEpsilon}
\bm{\varepsilon} = \frac{1}{2}(\nabla \bm{u} + \nabla \bm{u}^{\rm T}).
\end{equation}

In order to solve the evolution equation {for $\bm{\alpha}$} under the constraint \eqref{eq:divalpha}, {it is advantageous to} consider {a corresponding governing equation} on the plastic distortion $\uu{U}^{\rm p}$ {for which there are no constraints, except the periodicity.}  {From} equations \eqref{eq:EqFDM}$_3$ and \eqref{eq:defUp}, the plastic distortion is connected to the Nye tensor {by the relation}
\begin{equation}\label{eq:RelationAlphaUp}
\bm{\alpha} = - \uu{curl}(\uu{U}^{\rm p}).
\end{equation}
{Obviously, the plastic distortion is not uniquely defined by the Nye tensor. Relations \eqref{eq:EqFDM}$_4$
and \eqref{eq:RelationAlphaUp} imply that the rate of plastic distortion $\dot{\uu{U}}^{\rm p}$ is given by,
up to a constant second-order tensor,
\begin{equation}
\dot{\uu{U}}^{\rm p}= \bm{\alpha}\times \uu{V}+\nabla\bm{\phi}.
\end{equation}
{Since we are investigating dislocation-mediated plasticity, we assume that the rate of plastic distortion becomes nil when dislocations have zero velocity. Consequently we assume that $\nabla\bm{\phi}=\uu{0}$}\footnote{{It should be noted that this constitutive assumption has already been formulated by \cite{acharya_new_2010}.}}. The transport equation \eqref{eq:EqFDM}$_4$ can be rewritten in terms of the plastic distortion
\begin{equation}\label{eq:TransportEqUp}
\dot{\uu{U}}^{\rm p} =  - \uu{curl}(\uu{U}^{\rm p})  \times \uu{V}.
\end{equation}
{Solving this differential equation for $\uu{U}^{\rm p}$, with appropriate initial conditions,} one can then deduce the dislocation density tensor $\bm{\alpha}$ using equation \eqref{eq:RelationAlphaUp} which directly ensures the constraint \eqref{eq:divalpha}. It is noted that, given an initial dislocation density field $\bm{\alpha}_0$, an initial incompatible (i.e. gradient-free) plastic distortion $\uu{U}^{\rm p}_0$ can be determined by solving \aj{the Poisson equation}
\begin{equation}\label{eq:InitCond}
\bm{\Delta}\uu{U}^{\rm p}_0=\uu{curl}(\bm{\alpha}_0).
\end{equation}

{The FDM {problem \eqref{eq:EqFDM} thus reads alternatively}
\begin{align}\label{eq:EqFDM2}
\left\{\begin{array}{lll}
\uu{div}(\bm{\sigma}) & = & 0  \\
\bm{\sigma} & = & \uu{C}:(\nabla  \tou{u}-\uu{U}^{\rm p})  \\
\dot{\uu{U}}^{\rm p} & = & -  \uu{curl}(\uu{U}^{\rm p})  \times \uu{V} \\
\end{array}\right.
\end{align}
with initial conditions \eqref{eq:InitCond} for $\uu{U}^{\rm p}$. The following periodic boundary conditions are assumed
\begin{equation}\label{eq:bc}
 \tou{u} - \moy{\tod{\eps(u)}}.\tou{x} \;\text{periodic},\quad \tod{\sig}.\tou{n} \; \text{anti-periodic},
\end{equation}
where the total strain is given by equation \eqref{eq:defEpsilon}.} Macroscopic loadings are finally considered, expressed either in stress $\moy{\tod{\sig}}= \macr{\tod{\sig}}$ or strain $\moy{\tod{\eps}}= \macr{\tod{\eps}}$ {(or a combination of both)}, where $\macr{\tod{\sig}}$ and $\macr{\tod{\eps}}$ are respectively the prescribed macroscopic stress and strain and $\langle . \rangle$
denotes the spatial average over the unit-cell $V$.

It is interesting to note that in the present formulation of plasticity mediated by the motion of dislocations, the structure of the problem differs from that of classical (macroscopic) elastoplasticity by the constitutive relations expressed here by \eqref{eq:InitCond} and \eqref{eq:EqFDM2}$_3$, the other equations being preserved. Unlike in engineering plasticity, there is no explicit yield condition {on the stress $\tod{\sig}$} and the rate of plastic distortion does not derive from some normality property but directly arises from the motion of dislocations under applied stress. It therefore remains to specify how the dislocation velocity $\uu{V}$ depends on the other unknowns of the problem, in particular the stress field.

\subsection{Constitutive law for the dislocation's velocity}\label{sec:VelocityLaw}
Several studies of dislocation motion from molecular dynamics \citep{groh_dislocation_2009,ruestes_probing_2015,oren_dislocation_2017,cho_mobility_2017}  have shown that dislocation glide kinetics may be divided into three regimes: (i) an exponential dependence on the stress at velocities up to $10^{-3}C_T$, where $C_T$ is the transverse sound wave velocity, (ii) a linear stress-velocity relationship in the range of $10^{-3} - 10^{-1} C_T$ and (iii) an asymptotic behavior for high subsonic and transonic velocities. In \aj{most} previous works of dislocations dynamics \citep{zbib_plastic_1998} or field dislocation mechanics \citep{acharya_new_2010,zhang_single_2015}, only the linear regime was considered. \aj{(Note that in the context of the PMFDM, \cite{acharya-and-roy-2006,puri_mechanical_2011} considered also a power dependence of the velocity on stress).} {This} basically corresponds to the case of quasi-static plasticity at moderate stress levels. Consequently, we shall here also focus on the linear regime. The phenomenological law for the dislocation velocity is supposed to be of the form
\begin{equation}\label{eq:defV}
\uu{V} = \frac{\uu{F}}{\eta ||\bm{\alpha}||},
\end{equation}
where $\eta>0$ is a viscous drag coefficient (depending on the material considered) and $\uu{F}$ is \aj{a driving force to be defined}. This choice for the velocity law corresponds to the {\it nonlocal level set} model of \cite{zhang_single_2015}. \\

\aj{In the case of dislocation motion} with no lattice friction, the driving force is given by \citep{acharya_driving_2003}
\aj{
\begin{equation}\label{eq:defF}
\uu{F} = -\tod{\epsilon}:{(\tod{\sig}.\tod{\alpha})}.
\end{equation}
with $\tod{\epsilon}$  the permutation tensor (see  \ref{ap:PKF}). It should be noted that for a single dislocation, the driving force corresponds to the Peach-Koehler force of classical dislocation theory. Such law {\it alone} does not account for lattice friction and thus does not contain a Peierls-type threshold \citep{peierls_size_1940}: under any arbitrary stress, dislocation densities are automatically moving.} \\

\aj{In practice, the introduction of lattice friction is mandatory in dislocation-mediated plasticity in order to account for energetic barriers. The first approach to model lattice friction consists in the introduction of a Peierls-type threshold directly in the dislocation mobility law, as it is done classically in DDD simulations \citep{kubin_13,po_ghoniem_14}: if the stress is below the threshold there is no motion, and if it reaches the threshold, equation \eqref{eq:defV} applies. The second approach consists in the introduction of non-convex energy density functions in the mechanical dissipation (see \cite{zhang_single_2015,das_microstructure_2016}). This approach is closer to the physics since it permits to keep the memory of the discrete nature of dislocations and to access dislocations patterning. In the following, a non-convex energy density function is introduced, following the work of \cite{zhang_single_2015,das_microstructure_2016}. The volumic density of free (stored) energy $w$ is assumed to be of the form
\begin{equation}\label{eq:StoredEnergy}
w=\frac{1}{2} \bm{\epsilon}^e:\mathbf{C}:\bm{\epsilon}^e + G(\uu{U}^{\rm p}) 
\end{equation}
where $\bm{\epsilon}^e$ is the symmetric part of the elastic distortion $\uu{U}^{\rm e}$. The function $G$ is supposed to be multi-well non-convex which corresponds to an energy function with barriers to slip, thus enabling preferred energetic status to certain plastic strains. It is possible to add extra terms in equation \eqref{eq:StoredEnergy} to account for instance for the core energy, as done in \cite{zhang_single_2015}. This path was not followed here since the core energy term can lead to numerical issues in the resolution of the hyperbolic evolution equation which is known to be very sensitive to small perturbations \citep{leveque_finite_2002}. With the constitutive assumption \eqref{eq:StoredEnergy}, the study of the intrinsic dissipation (see \ref{ap:PKF}) leads to the definition of the driving force associated to the velocity field $\mathbf{V}$:
\begin{equation}
   \uu{F} = -\boldsymbol{\epsilon}:\left(\left(\bm{\sigma}- \frac{\partial G}{\partial \uu{U}^\abr{p}} \right).\boldsymbol{\alpha}
\right).
\end{equation}}

\section{Numerical integration of the constitutive model}\label{sec:numeric}
\subsection{{General resolution procedure}}
\subsubsection{Generalities}\label{sec:GenImpl}
The numerical integration of FDM equations consists in finding for {a given plastic distortion  $\uu{U}^{\rm p}$} and some boundary conditions, the total displacement $\bm{u}$ and stress $\bm{\sigma}$ solving system \eqref{eq:EqFDM2}. One of the main difficulties is to ensure simultaneously the {elastic} equilibrium  (elliptic equation) and the transport of dislocation (transport equation). The strategy adopted in this work is to treat {\it separately} the static problem and the {evolution problem (i.e. transport equation)} through an {alternating-directions} procedure. The reason for this choice is that the two systems require specific solvers which can hardly be used simultaneously.

{In practice, the resolution consists in finding the mechanical state  $\mathcal{S}_{n+1}= \left\{\bm{u}_{n+1}\right.$, $\bm{\sigma}_{n+1}$, $\left.\uu{U}^{\rm p}_{n+1}  \right\}$~at time $t_{n+1}$, knowing the previous mechanical state $\mathcal{S}_{n}= \left\{\bm{u}_{n}\right.$, $\bm{\sigma}_{n}$, $\left.\uu{U}^{\rm p}_{n}  \right\}$~at time $t_{n}$ and considering boundary conditions  \eqref{eq:bc}. The static problem is \aj{first} solved for the {\it previous} plastic distortion
\begin{align}\label{eq:staticdiscr}
\left\{\begin{array}{lll}
\uu{div}(\bm{\sigma}_{n+1}) & = & 0  \\
\bm{\sigma}_{n+1} & = & \uu{C}:(\nabla \bm{u}_{n+1}-\uu{U}^{\rm p}_{n}),
\end{array}\right.
\end{align}
then the plastic distortion is updated by solving the transport equation with the {\it new} stress field
\begin{equation}\label{eq:EqFDMEvolution}
\dot{\uu{U}}^{\rm p}_{n+1}  =  -  \uu{curl}(\uu{U}^{\rm p}_{n})  \times \uu{V}(\uu{U}^{\rm p}_{n},\bm{\sigma}_{n+1}).
\end{equation}
In the following, the subscripts $n$ and $n+1$ will be omitted to reduce the amount of notation.}

\subsubsection{Static problem}\label{eq:static_pb}
In order to solve the static problem \eqref{eq:staticdiscr}, we consider the FFT scheme proposed by \cite{brenner_numerical_2014} (see also \cite{berbenni_numerical_2014}), based on the work of \cite{moulinec_numerical_1998}.

Let us first consider the case of a {\it homogeneous} elastic medium with moduli tensor $\uu{C}(\uu{x}) = \uu{C}^0$; the solution field $\bm{\varepsilon}$ can be classically expressed as
\begin{equation}
\bm{\varepsilon}(\uu{x}) = \langle \bm{\varepsilon} \rangle - (\bm{\Gamma}^0 \ast {\bm{\tau}}) (\uu{x})\quad \quad {\rm where} \quad {\bm\tau} = - \uu{C}^0:{\uu{U}}^{\rm p}.
\end{equation}
In this equation, $\ast$ is the convolution product and $\bm{\Gamma}^0$ the Green operator of the homogeneous medium with elasticity $\uu{C}^0$.  In the Fourier space, this equation reads
\begin{equation}
\widehat{\bm{\varepsilon}}(\bm{\xi}) = - \widehat{\bm{\Gamma}}^0(\bm{\xi}) : \widehat{{\bm\tau}}(\bm{\xi}),\quad \quad \forall \bm{\xi}\neq \uu{0},
\end{equation}
where the Fourier transform of the Green operator $\bm{\Gamma}^0$ is recalled:
\begin{equation}
\widehat{\bm{\Gamma}}^0(\bm{\xi}) = \left[ \bm{\xi} \otimes \left(\bm{\xi}.\uu{C}^0.\bm{\xi} \right)^{-1} \otimes \bm{\xi}  \right]^{(s)}.
\end{equation}
In this equation, the symbol $[.]^{(s)}$ indicates minor and major symmetrization. {The Fourier transform of the stress field} reads
\begin{equation}\label{eq:HomogeneousSigmaFFT}
\widehat{\bm{\sigma}}(\bm{\xi}) =\left(\uu{C}^0:\widehat{\bm{\Gamma}}^0(\bm{\xi}) - \uu{I} \right) : \uu{C}^0: \widehat{\uu{U}}^{\rm p}(\bm{\xi}),\quad \quad \forall \bm{\xi}\neq \uu{0}, \quad \quad \widehat{\bm{\sigma}}(\bm{0}) = \macr{\bm{\sigma}},
\end{equation}
where $\uu{I}$ is the fourth-order identity tensor. {Besides, from the definition \eqref{eq:RelationAlphaUp},} the Fourier transform of the {incompatible} plastic distortion ${\uu{U}}^{\rm p}$ is given by \citep{brenner_numerical_2014}
\begin{equation}
\widehat{\uu{U}}^{\rm p}(\bm{\xi}) = \imath \frac{\widehat{\bm{\alpha}}(\bm{\xi})\times \bm{\xi}}{\| \bm{\xi}\|^2} ,\quad \quad \forall \bm{\xi}\neq \uu{0}, \quad \quad {\widehat{\uu{U}}^{\rm p}(\bm{0}) = \macr{\uu{U}}^{\rm p}},
\end{equation}
where $\imath$ is the imaginary unit. \\

In the case of a {\it heterogeneous} elastic medium with moduli tensor $\uu{C}(\uu{x})$, the local behavior can be rewritten as
\begin{equation}
\bm{\sigma} = \uu{C}^0:\nabla \bm{u} + \bm{\tau}
\end{equation}
where
\begin{equation}
 \bm{\tau} = - \uu{C}:\uu{U}^{\rm p} + (\uu{C}-\uu{C}^0):\nabla \bm{u}{,}
 \end{equation}
{with the uniform moduli $\uu{C}^0$ of a reference medium.} The only difference with the homogeneous elastic problem is that the prescribed eigentress field $\bm{\tau}$ is not known {a priori} since it depends on the field  $\bm{u}$ which solves the problem. When the reference medium is adequately chosen, the solution field $\bm{\varepsilon}$ is obtained as a series expansion:
\begin{equation}
\bm{\varepsilon}(\uu{x}) = \sum_{i=0}^{+\infty} \left( - \bm{\Gamma}^0 \ast \delta \uu{C} (\uu{x}) \right)^i:\left( \langle \bm{\varepsilon} \rangle + (\bm{\Gamma}^0 \ast \uu{C}:\uu{U}^{\rm p}) (\uu{x}) \right).
\end{equation}
Efficient iterative numerical procedures based on fast-Fourier tranforms (FFT) may then be used to compute the solution field $\bm{\varepsilon}$ {\citep[see, among others,][] {moulinec_numerical_1998,brisard_2012,moulinec_comparison_2014,schneider_2019}.}

\subsubsection{Evolution problem}
As explained {above}, the evolution problem \eqref{eq:EqFDMEvolution} is treated separately from the static problem. {In order to emphasize the main characteristics of the constitutive transport problem, it is useful to detail the full set of equations.} Knowing the stress state, the evolution problem \eqref{eq:EqFDMEvolution} consists of a system of nine hyperbolic equations
\begin{align}\label{eq:BC}
\left\{\begin{array}{lll}
\dot{U}^{\rm p}_{11} & = & (U^{\rm p}_{12,1}-U^{\rm p}_{11,2})V_{2} - (U^{\rm p}_{11,3}-U^{\rm p}_{13,1})V_{3} \\
\dot{U}^{\rm p}_{12} & = & (U^{\rm p}_{13,2}-U^{\rm p}_{12,3})V_{3} - (U^{\rm p}_{12,1}-U^{\rm p}_{11,2})V_{1} \\
\dot{U}^{\rm p}_{13} & = & (U^{\rm p}_{11,3}-U^{\rm p}_{13,1})V_{1} - (U^{\rm p}_{13,2}-U^{\rm p}_{12,3})V_{2} \\
\dot{U}^{\rm p}_{21} & = & (U^{\rm p}_{22,1}-U^{\rm p}_{21,2})V_{2} - (U^{\rm p}_{21,3}-U^{\rm p}_{23,1})V_{3} \\
\dot{U}^{\rm p}_{22} & = & (U^{\rm p}_{23,2}-U^{\rm p}_{22,3})V_{3} - (U^{\rm p}_{22,1}-U^{\rm p}_{21,2})V_{1} \\
\dot{U}^{\rm p}_{23} & = & (U^{\rm p}_{21,3}-U^{\rm p}_{23,1})V_{1} - (U^{\rm p}_{23,2}-U^{\rm p}_{22,3})V_{2} \\
\dot{U}^{\rm p}_{31} & = & (U^{\rm p}_{32,1}-U^{\rm p}_{31,2})V_{2} - (U^{\rm p}_{31,3}-U^{\rm p}_{33,1})V_{3} \\
\dot{U}^{\rm p}_{32} & = & (U^{\rm p}_{33,2}-U^{\rm p}_{32,3})V_{3} - (U^{\rm p}_{32,1}-U^{\rm p}_{31,2})V_{1} \\
\dot{U}^{\rm p}_{33} & = & (U^{\rm p}_{31,3}-U^{\rm p}_{33,1})V_{1} - (U^{\rm p}_{33,2}-U^{\rm p}_{32,2})V_{2}. \\
\end{array}\right.
\end{align}
From the definition \eqref{eq:defV} of the velocity, it appears that the evolution problem consists in a {\it vectorial}, {\it multi-dimensional} and {\it non-linear} hyperbolic system of Hamilton-Jacobi type. Several approaches permit to solve Hamilton-Jacobi equations: ENO and WENO schemes \citep{osher_high-order_1991,jiang_weighted_2000}, discontinuous Galerkin finite element \citep{hu_discontinuous_1999} and Godunov-type approaches \citep{lin_high-resolution_2000,kurganov_semidiscrete_2001}. These approaches rely on advanced numerical methods that are unfortunately restricted{, up to now,} to two-dimensional scalar problems; {if the extension to the three-dimensional case does not seem to be an unrealistic task, the extension to vectorial equations remains an open and difficult problem.} \\

As a first step towards the resolution of the full coupled problem of FDM, we first consider a {particular} case  {for which} the vectorial system {reduces} to a scalar equation. {This allows us to resort to}  efficient numerical solvers for Hamilton-Jacobi equations.

\aj{
\subsection{A simplified layer problem}\label{eq:simplifedlayer}
\subsubsection{Position of the problem}
We consider a simplified model problem of edge and screw dislocations confined in a thin layer as shown in Figure \ref{fig:LayerProblem}. This model problem can be seen as an elastoplastic body where plastic flow is constrained in a layer, acting as the slip plane. Thus, in the layer, both edge and screw dislocations exist and FDM is active, while the two outer regions are purely elastic linear. This model may be viewed as an extension to three dimensions of the problem considered by \cite{zhang_single_2015}.} \\

\begin{figure}[!ht]
\centering
\includegraphics[width=12cm]{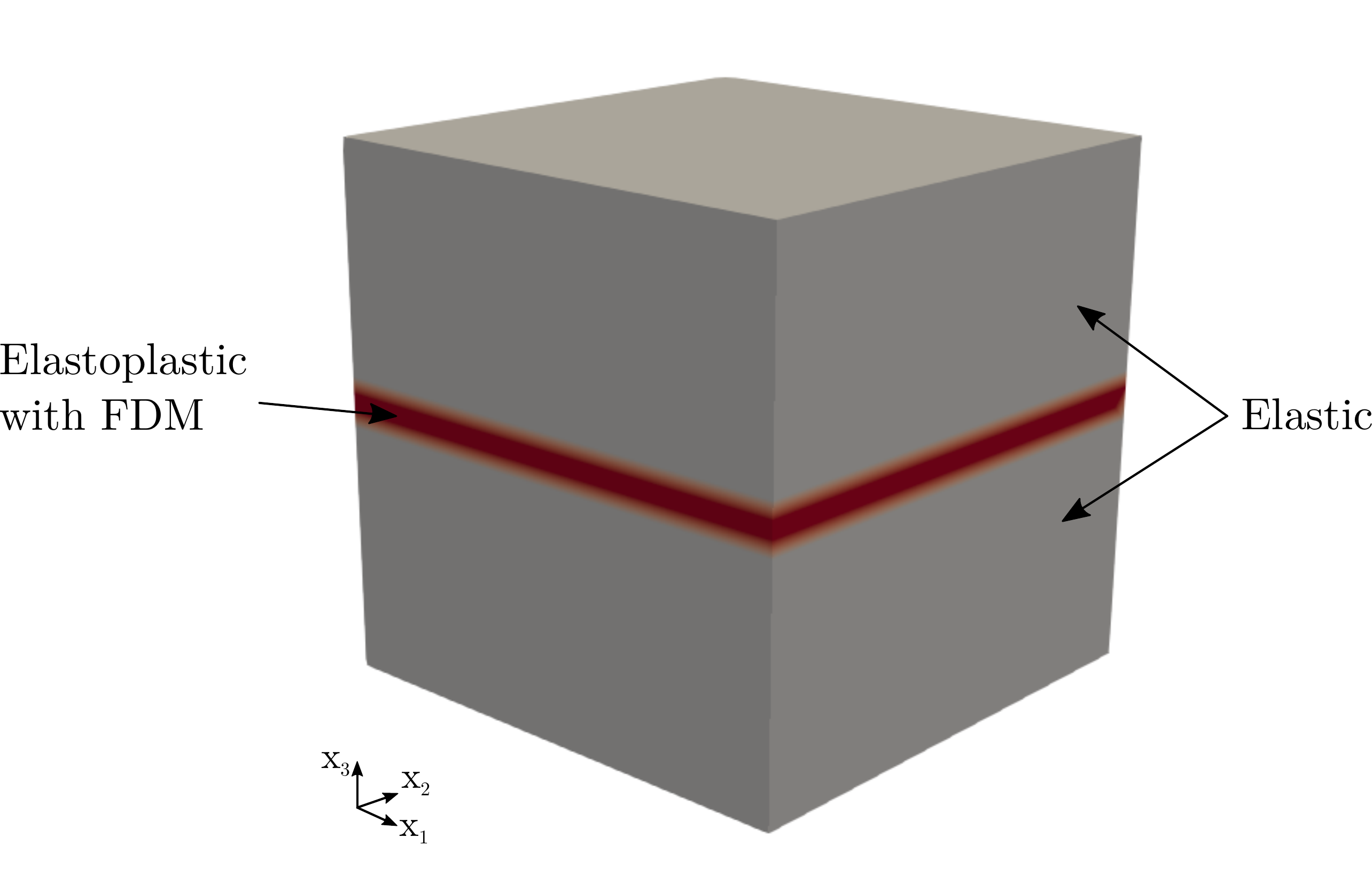}
\caption{Simplified layer problem.}
\label{fig:LayerProblem}
\end{figure}

\aj{In the layer, the following ansatz is assumed:
\begin{enumerate}
\item The plastic distortion is supposed to be constrained in the layer and is of the form
\begin{equation}
\uu{U}^{\rm p} = U_{13}^{\rm p} (x_1,x_2) \uu{e}_{1}\otimes\uu{e}_{3}.
\end{equation}
This implies that the dislocation density tensor is of the form
\begin{equation}
\bm{\alpha} = \alpha_{11}(x_1,x_2) \uu{e}_{1}\otimes\uu{e}_{1} + \alpha_{12}(x_1,x_2) \uu{e}_{1}\otimes\uu{e}_{2},
\end{equation}
where $\alpha_{11}(x_1,x_2)=-U_{13,2}^{\rm p}$ and $\alpha_{12}(x_1,x_2)=U_{13,1}^{\rm p}$. 
\item Let us suppose that the non-convex energy function $G$ is of the form
\begin{equation}
G =  \beta {\rm sin}\left(  \frac{U_{13}^{\rm p}}{\beta}\right) \tau_y 
\end{equation}
where $\beta$ is a nondimensional parameter that is supposed to be very small \citep{das_microstructure_2016} and $\tau_y$ has the dimension of a stress. The term $\partial G / \partial U_{13}^{\rm p}$ arising in the definition of the driving force is thus a high-frequency oscillatory function bounded with an amplitude $\tau_y$. This energy, which was referred as a wiggly energy \citep{das_microstructure_2016}, permits to account for lattice friction produced by a discrete lattice, with a threshold in stress corresponding to $\tau_y$.
\item Dislocations move in the (thin) layer where they are constrained to stay (i.e. no dislocation climb or cross-slip) so the dislocation velocity $\uu{V}$ can be assumed  in the form
\begin{equation}
\uu{V} = V_1(x_1,x_2)\uu{e}_{1}+V_2(x_1,x_2)\uu{e}_{2}.
\end{equation}
\end{enumerate}
The dissipation $\mathcal{D}$ can be written as (see \ref{ap:PKF})
\begin{align}
\mathcal{D} &=  \int_{\rm Layer} \left(\bm{\sigma}- \frac{\partial G}{\partial \uu{U}^\abr{p}} \right):\left(\boldsymbol{\alpha}\times\mathbf{V}\right)
\dd v= \int_{\rm Layer} \left(\sigma_{13} - \frac{\partial G}{\partial U_{13}^{\rm p}} \right) (\alpha_{11}V_2 - \alpha_{12} V_1) \dd v \nonumber \\
 & = h \int_{\rm -L}^L  \int_{\rm -L}^L  \left(\tau_{13}- \frac{\partial G}{\partial U_{13}^{\rm p}} \right)(\alpha_{11}V_2 - \alpha_{12} V_1) \dd x_1 \dd x_2,
\end{align}
where $h$ is the layer's thickness and $\tau_{13}$ is the average stress in the layer given by
\begin{equation}
\tau_{13} = \frac{1}{h} \int_{-h/2}^{h/2} \sigma_{13} \dd x_3.
\end{equation}
Here, $\tau_{13}$, $\alpha_{11}$, $\alpha_{12}$, $V_1$, $V_2$ and $\partial G / \partial U_{13}^{\rm p}$ are functions of $(x_1,x_2)$ only. 
\begin{itemize}
\item[(iv)] The driving force is finally supposed to depend on the average stress $\tau_{13}$:
\begin{align}
\left\{\begin{array}{ll}
F_1 & = \ds -\left(\tau_{13}- \frac{\partial G}{\partial U_{13}^{\rm p}} \right) \alpha_{12} \\
F_2 & = \ds \left(\tau_{13}- \frac{\partial G}{\partial U_{13}^{\rm p}} \right) \alpha_{11}.
\end{array}\right.
\end{align}
This implies that the driving force $\uu{F}$ is a function of $(x_1,x_2)$ only, which permits to apply the constitutive law $\uu{V}=\uu{V}(\uu{F})$ in two dimensions. \\
\end{itemize}}

\aj{The problem {which has to be solved for $({\uu{u}},{\bm{\sigma}})$} reduces to
\begin{align}\label{eq:EqFDMelasto}
\left\{\begin{array}{lll}
&\ds \frac{\partial {\sigma}_{ij}}{\partial {x}_i}    & =  0 \\[0.35cm]
& {\sigma}_{ij}  &=  \ds \mu \left( \frac{\partial {u}_{i}}{\partial {x}_j} + \frac{\partial {u}_{j}}{\partial {x}_i} - U_{ij}^{\rm p} \right) + \lambda\frac{\partial {u}_{k}}{\partial {x}_k} \delta_{ij} \\[0.35cm]
& \ds \frac{\partial {U}_{13}^{\rm p}}{\partial {t}} &=  \ds \frac{\ds {\tau}_{13}-{\rm cos}\left(  \frac{U_{13}^{\rm p}}{\beta}\right) {\tau}_y}{{\eta}} \sqrt{ \left(\frac{\partial {U}_{13}^{\rm p}}{\partial {x}_1}\right)^2 + \left(\frac{\partial  {U}_{13}^{\rm p}}{\partial {x}_2}\right)^2 }\\[0.35cm]
& &  {\text{+ periodicity conditions},}
\end{array}\right.
\end{align}
with prescribed macroscopic strain $\ds \bar{\varepsilon}_{ij} = \frac{1}{2} \left\langle \frac{\partial {u}_{i}}{\partial {x}_j} + \frac{\partial {u}_{j}}{\partial {x}_i}\right\rangle$.} \\

\aj{Following \cite{zhang_single_2015}, a dimensional analysis suggests the introduction of dimensionless variables
\begin{equation}
\tilde{\uu{x}} = \frac{\uu{x}}{b},\quad \tilde{t}=\frac{V_{\rm s}t}{b},\quad \tilde{\uu{u}} = \frac{\uu{u}}{b},\quad \tilde{\bm{\sigma}} = \frac{\bm{\sigma}}{\mu},\quad \tilde{\tau}_{13} = \frac{\tau_{13}}{\mu} ,\quad \tilde{\tau}_{y} = \frac{\tau_{y}}{\mu} ,\quad\tilde{\eta} = \frac{V_{\rm s} \eta}{\mu},\quad \tilde{\bm{\alpha}} = b\bm{\alpha},
\end{equation}
where $b$ is the norm of the Burgers vector, $\mu$ is the elastic shear modulus and $V_{\rm s}=\sqrt{\mu/\rho}$ is the elastic shear wave speed ($\rho$ being the density).} \aj{The problem {which has to be solved for $(\tilde{\uu{u}},\tilde{\bm{\sigma}})$} thus reads
\begin{align}\label{eq:EqFDMelasto}
\left\{\begin{array}{lll}
&\ds \frac{\partial \tilde{\sigma}_{ij}}{\partial \tilde{x}_i}    & =  0 \\[0.35cm]
& \tilde{\sigma}_{ij}  &=  \ds \left( \frac{\partial \tilde{u}_{i}}{\partial \tilde{x}_j} + \frac{\partial \tilde{u}_{j}}{\partial \tilde{x}_i} - U_{ij}^{\rm p} \right) + \frac{\lambda}{\mu}\frac{\partial \tilde{u}_{k}}{\partial \tilde{x}_k} \delta_{ij} \\[0.35cm]
& \ds \frac{\partial {U}_{13}^{\rm p}}{\partial \tilde{t}} &=  \ds \frac{\ds \tilde{\tau}_{13}-{\rm cos}\left(  \frac{U_{13}^{\rm p}}{\beta}\right) \tilde{\tau}_y}{\tilde{\eta}} \sqrt{ \left(\frac{\partial {U}_{13}^{\rm p}}{\partial \tilde{x}_1}\right)^2 + \left(\frac{\partial  {U}_{13}^{\rm p}}{\partial \tilde{x}_2}\right)^2 }\\[0.35cm]
& &  {\text{+ periodicity conditions},}
\end{array}\right.
\end{align}
with prescribed macroscopic strain $\ds \bar{\varepsilon}_{ij} = \frac{1}{2} \left\langle \frac{\partial \tilde{u}_{i}}{\partial \tilde{x}_j} + \frac{\partial \tilde{u}_{j}}{\partial \tilde{x}_i}\right\rangle$.} \\

\aj{This model problem is interesting from a computational point of view because it allows to express the evolution problem in two dimensions only without losing too much of the physics. Indeed, the problem is 3D in Fourier space and only the transport equation is constrained in the layer, which ultimately corresponds to impose the slip plane. The static problem, corresponding to equations \eqref{eq:EqFDMelasto}$_{1-2}$, can easily be solved in Fourier space using the FFT scheme described in Section \ref{eq:static_pb}. Once the shear stress $\sigma_{13}$ is computed, it is easy to compute the average stress $\tau_{13}$ and, as explained in Section \ref{sec:GenImpl}, ${U}_{13}^{\rm p}$ is updated by solving the evolution problem, whose algorithm is presented hereafter.}

\subsubsection{Resolution of the evolution problem - \aj{1D case}}
The evolution problem consists of an hyperbolic Hamilton-Jacobi equation {\eqref{eq:EqFDMelasto}$_{3}$}. It is well known that such equations requires specific solvers in order to avoid spurious numerical effects as oscillations and damping   \citep{leveque_finite_2002}.

{Following the suggestion of \cite{das_microstructure_2016}, we adopt here} \cite{kurganov_semidiscrete_2001}'s scheme which {is a Godunov-type high resolution scheme. As such, it} combines simplicity and accuracy. \aj{First t}he algorithm is presented in one-dimension. We thus consider the following one-dimensional prototype equation
\begin{equation}\label{eq:HJ1D}
\frac{\partial \phi}{\partial \tilde{t}} + H\left(\frac{\partial \phi}{\partial x}\right) =0.
\end{equation}
where $\phi={U}_{13}^{\rm p}$ and $x=\tilde{x}_1$. \aj{The Hamiltonian $H$ reads
\begin{equation}
H\left(\frac{\partial \phi}{\partial x}\right) = \frac{\ds {\rm cos}\left(  \frac{\phi}{\beta}\right) \tilde{\tau}_y-\tilde{\tau}_{13}}{\tilde{\eta}} \left|\frac{\partial \phi}{\partial x}\right|=v_0\left|\frac{\partial \phi}{\partial x}\right|,
\end{equation}
where the term $v_0$ is given by the previous time step.} \\

{A} uniform {resolution} grid {is chosen} and we use the following notations: $x_{j} = j \Delta x$ \aj{(corresponding to the nodes of the pixels introduced in the FFT algorithm)}, ${t}^n = n \Delta {t}$ and $\phi_j^n = {U}_{13}^{\rm p}(x_{j},t^n)$, where $\Delta x$ and $\Delta {t}$ are respectively the spatial scale and the time step. Assuming that the point values of ${U}_{13}^{\rm p}$ at time $\tilde{t} = {t}^n$ {$\left(\phi_j^{n}\right)$} {are known}, we are looking for the point values of ${U}_{13}^{\rm p}$ at time $\tilde{t} = {t}^{n+1}$ {$\left(\phi_j^{n+1}\right)$}.

\paragraph*{Step 1: construction of a continuous piecewise interpolant.} We start with the construction of the continuous piecewise interpolant $\breve{\phi}(x,t^n)$  in order to avoid spurious oscillations. The quadratic interpolant over the interval $[x_j,x_{j+1}]$ reads
\begin{equation}\label{eq:interpolant}
\breve{\phi}(x,t^n) = \phi_j^n + \frac{(\Delta \phi)_{j+1/2}^n}{\Delta x}(x-x_j) + \frac{(\Delta \phi)'_{j+1/2}}{2(\Delta x)^2}(x-x_j)(x-x_{j+1}),
\end{equation}
where
\begin{equation}
\Delta \phi_{j+1/2}^n = \phi_{j+1}^n - \phi_j^n.
\end{equation}
The term ${(\Delta \phi)'_{j+1/2}}/{(\Delta x)^2}$ is an approximation of the second derivative $\phi_{xx}(x_{j+1/2},t^n)$. A nonlinear limiter is used to compute this derivative in order to ensure the nonoscillatory nature of $\breve{\phi}(x,t^n)$. {A} one-parameter family of the minmod limiters {is used}
\begin{align}
(\Delta \phi)'_{j+1/2} = {\rm minmod}& \left(  \theta \left[ (\Delta \phi)_{j+3/2}^n - (\Delta \phi)_{j+1/2}^n \right], \frac{1}{2} \left[ (\Delta \phi)_{j+3/2}^n -  (\Delta \phi)_{j-1/2}^n \right] \right. \nonumber\\
& \left. , \theta \left[ (\Delta \phi)_{j+1/2}^n -  (\Delta \phi)_{j-1/2}^n\right] \right),
\end{align}
where $\theta \in [1,2]$ and the minmod function is defined by
\begin{equation}\label{eq:flowrule}
{\rm minmod}(x_1,x_2,...) = \left\{ \begin{array}{llll}
{\rm min}_j\{x_j\} & \quad & {\rm if}~x_j>0~\forall j, \\
{\rm max}_j\{x_j\} & \quad & {\rm if}~x_j<0~\forall j, \\
0 & \quad & {\rm otherwise}. \\
                       \end{array}
                     \right.
\end{equation}

\paragraph*{Step 2: estimation of the one-sided local speed of propagation.} We estimate the one-sided speed of propagation at the grid point $x_j$, which are given by
\begin{equation}
a_j^+ = \max\{H'(\phi_x^+),H'(\phi_x^-),0\};\quad \quad a_j^- = \min\{H'(\phi_x^+),H'(\phi_x^-),0\},
\end{equation}
where $\phi_x^\pm = \breve{\phi}_x(x_j \pm 0,t^n)$. Using the  continuous piecewise quadratic polynomial \eqref{eq:interpolant}, one gets
\begin{equation}
\phi_x^\pm = \frac{(\Delta \phi)_{j \pm 1/2}^n}{\Delta x} \mp \frac{(\Delta \phi)'_{j \pm 1/2}}{2\Delta x}.
\end{equation}

\paragraph*{Step 3: approximate solution of the Hamilton-Jacobi equation at intermediate grid points.} The Hamilton-Jacobi equation \eqref{eq:HJ1D} is exactly solved at intermediate points defined as $x_{j \pm} = x_{j}+a_j^{\pm} \Delta t$:
\begin{equation}
\phi^{n+1}_{j\pm} = \breve{\phi}(x_{j\pm},t^n) - \int_{t^n}^{t^{n+1}} H\left(\breve{\phi}_x(x_{j\pm},t)\right) {\rm d}t.
\end{equation}

With an appropriate CFL number condition (see \cite{kurganov_semidiscrete_2001}), the integral on the right-hand side can be evaluated within second-order accuracy by the midpoint rule; this yields to the following approximate Riemann solver
\begin{equation}
\phi^{n+1}_{j\pm} = \breve{\phi}(x_{j\pm},t^n) - \Delta t
 H\left(\breve{\phi}_x(x_{j\pm}^n,t^n)\right).
\end{equation}

\paragraph*{Step 4: projection of the intermediate solution onto the original grid.} The solution previously obtained at the intermediate points $x_{j \pm}$ is projected onto the original grid
\begin{equation}
\phi_j^{n+1} = \frac{a_j^+}{a_j^+-a_j^-} \phi^{n+1}_{j-} - \frac{a_j^-}{a_j^+-a_j^-} \phi^{n+1}_{j+},
\end{equation}
which leads to the fully discrete scheme
\begin{equation}
\phi_j^{n+1} = \frac{a_j^+}{a_j^+-a_j^-} \left(\breve{\phi}(x_{j-},t^n) - \Delta t
 H\left(\breve{\phi}_x(x_{j-}^n,t^n)\right) \right)  - \frac{a_j^-}{a_j^+-a_j^-} \left(\breve{\phi}(x_{j+},t^n)) - \Delta t
 H\left(\breve{\phi}_x(x_{j+}^n,t^n)\right) \right).
\end{equation}

\subsubsection{\aj{Resolution of the evolution problem - 2D case}} \aj{We continue with the resolution of the evolution problem in the two-dimensional case. We consider the prototype equation
\begin{equation}\label{eq:HJ2D_reso}
\frac{\partial \phi}{\partial \tilde{t}} + H\left(\frac{\partial \phi}{\partial x},\frac{\partial \phi}{\partial y}\right) =0,
\end{equation}
where $\phi={U}_{13}^{\rm p}$, $x=\tilde{x}_1$ and $y=\tilde{x}_2$. The Hamiltonian $H$ reads
\begin{equation}
H\left(\frac{\partial \phi}{\partial x},\frac{\partial \phi}{\partial y}\right) = \frac{\ds {\rm cos}\left(  \frac{\phi}{\beta}\right) \tilde{\tau}_y-\tilde{\tau}_{13}}{\tilde{\eta}}  \sqrt{  \left( \frac{ \partial\phi }{\partial x} \right)^2 + \left( \frac{\partial \phi }{\partial y} \right)^2 } = v_0 \sqrt{  \left( \frac{ \partial\phi }{\partial x} \right)^2 + \left( \frac{\partial \phi }{\partial y} \right)^2 },
\end{equation}
where the term $v_0$ is given by the previous time step.} \\

\aj{Again, {a} uniform {resolution} grid (corresponding to the pixels' nodes) {is chosen} with the following notations: $x_{j} = j \Delta x$, $y_{k} = k \Delta y$, $t^n = n \Delta t$ and $\phi_{jk}^n = {U}_{13}^{\rm p}(x_{j},y_{k},t^n)$, where $\Delta x$, $\Delta y$ and $\Delta t$ are respectively the spatial scales and the time step. Assuming that the point values of ${U}_{13}^{\rm p}$ at time $\tilde{t} = t^n$ {$\left(\phi_{jk}^n\right)$} {are known}, we are looking for the point values of ${U}_{13}^{\rm p}$ at time $\tilde{t} = t^{n+1}$ {$\left(\phi_{jk}^{n+1}\right)$}. \\
\paragraph*{Step 1: construction of a continuous piecewise interpolant.} We start with the construction of the continuous piecewise interpolant $\breve{\phi}(x,y,t^n)$  in order to avoid spurious oscillations. The extension of the quadratic interpolant written in the one-dimensional case  \eqref{eq:interpolant} to the two-dimensional case is tedious but straightforward (see \cite{kurganov_new_2000} for the full detail).
\paragraph*{Step 2: estimation of the one-sided local speed of propagation.} Then we evaluate the one-sided local speeds of propagation in the $x-$ and $y-$directions. These values at the grid point $(x_j,y_k)$ are given by
\begin{align}
a_{jk}^+ = \displaystyle\max_{\pm}\left\{\frac{\partial H}{\partial \phi_x}(\phi_x^\pm,\phi_y^\pm),0\right\},\quad \quad a_{jk}^- = \displaystyle\min_{\pm}\left\{\frac{\partial H}{\partial \phi_x}(\phi_x^\pm,\phi_y^\pm),0\right\}, \nonumber\\
b_{jk}^+ = \displaystyle\max_{\pm}\left\{\frac{\partial H}{\partial \phi_y}(\phi_x^\pm,\phi_y^\pm),0\right\},\quad \quad b_{jk}^- = \displaystyle\min_{\pm}\left\{\frac{\partial H}{\partial \phi_y}(\phi_x^\pm,\phi_y^\pm),0\right\},
\end{align}
where $\phi_x^\pm = \breve{\phi}_x(x_j \pm 0,y_k,t^n)$ and $\phi_y^\pm = \breve{\phi}_y(x_j,y_k \pm 0,t^n)$ are the right and the left derivatives in the $x-$ and $y-$direction, deduced from the two-dimensional interpolant \citep{kurganov_new_2000}.
\paragraph*{Step 3: approximate solution of the Hamilton-Jacobi equation at intermediate grid points.} The Hamilton-Jacobi equation \eqref{eq:HJ2D_reso} is then solved at the intermediate points $(x_{j \pm}^n = x_{j}+a_{jk}^{\pm} \Delta t,~y_{k \pm}^n = y_{k}+b_{jk}^{\pm} \Delta t$). This leads to the approximate Riemann solver
\begin{equation}
\phi^{n+1}_{j\pm,k\pm} = \breve{\phi}( x_{j \pm}^n,y_{k \pm}^n ,t^n) - \Delta t
 H\left( \breve{\phi}_x(x_{j \pm}^n,y_{k \pm}^n ,t^n),\breve{\phi}_y(x_{j \pm}^n,y_{k \pm}^n ,t^n)\right).
\end{equation}
\paragraph*{Step 4: projection of the intermediate solution onto the original grid.} The solution previously obtained at the intermediate points $x_{j \pm}$ and $y_{k \pm}$ is projected onto the original grid which leads to the fully discrete scheme
\begin{align}
\phi^{n+1}_{jk}  = & & \frac{a_{jk}^-b_{jk}^-}{
(a_{jk}^+-a_{jk}^-)(b_{jk}^+-b_{jk}^-)} \left( \breve{\phi}( x_{j+}^n,y_{k+}^n ,t^n) - \Delta t
 H\left( \breve{\phi}_x(x_{j+}^n,y_{k+}^n ,t^n),\breve{\phi}_y(x_{j+}^n,y_{k+}^n ,t^n)\right) \right) \nonumber \\
 &  + & \frac{a_{jk}^-b_{jk}^+}{
(a_{jk}^+-a_{jk}^-)(b_{jk}^+-b_{jk}^-)} \left( \breve{\phi}( x_{j+}^n,y_{k-}^n ,t^n) - \Delta t
 H\left( \breve{\phi}_x(x_{j+}^n,y_{k-}^n ,t^n),\breve{\phi}_y(x_{j+}^n,y_{k-}^n ,t^n)\right) \right) \nonumber \\
 &  + & \frac{a_{jk}^+b_{jk}^-}{
(a_{jk}^+-a_{jk}^-)(b_{jk}^+-b_{jk}^-)} \left( \breve{\phi}( x_{j-}^n,y_{k+}^n ,t^n) - \Delta t
 H\left( \breve{\phi}_x(x_{j-}^n,y_{k+}^n ,t^n),\breve{\phi}_y(x_{j-}^n,y_{k+}^n ,t^n)\right) \right) \nonumber \\
 & + & \frac{a_{jk}^+b_{jk}^+}{
(a_{jk}^+-a_{jk}^-)(b_{jk}^+-b_{jk}^-)} \left( \breve{\phi}( x_{j-}^n,y_{k-}^n ,t^n) - \Delta t
 H\left( \breve{\phi}_x(x_{j-}^n,y_{k-}^n ,t^n),\breve{\phi}_y(x_{j-}^n,y_{k-}^n ,t^n)\right) \right).
\end{align}
}

\newpage
\section{Numerical results: uncoupled problems}\label{sec:results_uncoupled}

\aj{\subsection{Preliminaries}\label{sec:4.1}}
The aim of this section is  to study numerically the evolution of {the pointwise dislocation density tensor} by considering only the transport problem. \aj{To do so, we assume a prescribed constant stress field (in time and space) and a null stress threshold ($\tilde{\tau}_y=0$). In this way, we have to solve the hyperbolic equation
\begin{equation}\label{eq:HJ2D}
\frac{\partial {U}_{13}^{\rm p}}{\partial \tilde{t}} + v_0 \sqrt{ \left(\frac{\partial {U}_{13}^{\rm p}}{\partial \tilde{x}_1}\right)^2 + \left(\frac{\partial  {U}_{13}^{\rm p}}{\partial \tilde{x}_2}\right)^2 } = 0,
\end{equation}
where the ``celerity'' of dislocation $v_0=-\tilde{\tau}_{13}/\tilde{\eta}$ is constant since we do not consider the coupling with the static problem.}

This {first step} is needed in order to assess solely the algorithm proposed for the hyperbolic Hamilton-Jacobi system because in the particular case of equation \eqref{eq:HJ2D}, analytical solutions and mathematical properties \aj{may} be exhibited. 
%
%
Indeed, in terms of the dislocation densities $\alpha_{11}$ and $\alpha_{12}$, the hyperbolic equation \eqref{eq:HJ2D} reads
\begin{align}\label{eq:A1A2Dot}
\left\{\begin{array}{lll}
\dot{\alpha}_{11} & = & - (\alpha_{11}V_2 - \alpha_{12}V_{1})_{,2} \\
\dot{\alpha}_{12} & = & - (\alpha_{12}V_1 - \alpha_{11}V_{2})_{,1}
\end{array}\right.
\end{align}
where the velocities $V_1$ and $V_2$ are given by
\begin{align}\label{eq:V1V2}
\left\{\begin{array}{lll}
V_1 & = & \ds -\frac{\ds\tilde{\tau}_{13}}{\tilde{\eta}}  \ds \frac{\alpha_{12}}{\sqrt{\alpha_{11}^2+\alpha_{12}^2}} = v_0 \ds \frac{\alpha_{12}}{\sqrt{\alpha_{11}^2+\alpha_{12}^2}} \\[0.5cm]
V_2 & = & \ds \frac{\tilde{\tau}_{13}}{\tilde{\eta}}  \ds \frac{\alpha_{11}}{\sqrt{\alpha_{11}^2+\alpha_{12}^2}} = -v_0 \ds \frac{\alpha_{11}}{\sqrt{\alpha_{11}^2+\alpha_{12}^2}} .
\end{array}\right.
\end{align}
Then, let us study the derivative of \aj{$\|{\alpha}\|  = \sqrt{\alpha_{11}^2+\alpha_{12}^2}$} in the velocity $\uu{V}$ defined as
\begin{equation}\label{eq:MatDerivativeA}
\frac{{\rm d}\|{\alpha}\|}{{\rm d} t} = \dot{\|{\alpha}\|} + {\uu{V}.{\bm{\nabla}{\|{\alpha}\|}}},
\end{equation}
where
\begin{align}
\left\{\begin{array}{lll}
\dot{\|{\alpha}\|} & = &  \ds \frac{{1}}{{\|{\alpha}\|}} (\dot{\alpha}_{11}\alpha_{11}+\dot{\alpha}_{12}\alpha_{12}) \\[0.5cm]
{\uu{V}.{\bm{\nabla}{\|{\alpha}\|}}} & = & \ds \frac{{1}}{{\|{\alpha}\|}} \Big[ (\alpha_{11}\alpha_{11,1} + \alpha_{12}\alpha_{12,1})V_1 + (\alpha_{11}\alpha_{11,2}  + \alpha_{12}\alpha_{12,2} )V_2 \Big].
\end{array}\right.
\end{align}
According to the definition of $V_1$ and $V_2$ given by \eqref{eq:V1V2}, the transport equation \eqref{eq:A1A2Dot} reduces to
\begin{align}
\left\{\begin{array}{lll}
\dot{\alpha}_{11} & = & - \alpha_{11,2}V_2 + \alpha_{12,2}V_{1} \\[0.2cm]
\dot{\alpha}_{12} & = & \alpha_{11,1}V_{2} - \alpha_{12,1}V_1.
\end{array}\right.
\end{align}
\aj{Thus the derivative of $\|{\alpha}\|$ \eqref{eq:MatDerivativeA} in the velocity field $\uu{V}$ reads}
\begin{equation}\label{eq:conservalpha}
\frac{{\rm d}\|{\alpha}\|}{{\rm d} t} = {\ds \frac{{\alpha_{11}V_1 + \alpha_{12}V_2}}{{\|{\alpha}\|}} \ds\left( \alpha_{11,1}  + \alpha_{12,2}\right)=0}
\end{equation}
due to the relation \eqref{eq:divalpha}. \aj{This means that the transport of $\|{\alpha}\|$ modeled by equation \eqref{eq:HJ2D} is thus {\it conservative} in the velocity field considered\footnote{{It should be noted that, if annihilation occurs, the velocity is \aj{no} longer well defined at the shock front, and thus equation \eqref{eq:conservalpha} is no longer valid.}}: no damping and no spreading of $\|{\alpha}\|$ should be observed in the numerical simulations.  It should be noted that this property only holds for the constitutive law \eqref{eq:defV} and under the hypothesis of constant and uniform velocity $v_0$. This provides however, under these hypotheses, a valuable assessment to test the accuracy of the time-integration algorithm for the transport equation. An explicit Euler algorithm would not satisfy this condition.} \\

\aj{\subsection{Description of the simulations}\label{sec:4.2}}

{In the sequel, the FDM approach is used for two model problems with single dislocations lines or loops, namely (i) the annihilation of edge dislocations and (ii) the expansion of dislocation loops.}
To perform the simulations, dislocation densities and material parameters need to be prescribed. We consider \aj{only} a 2D unit-cell domain (since we do not solve the static problem) of $320b \times 320b$ \aj{corresponding to the layer}, with $b$ the norm of {the chosen} Burgers vector {$\uu{b}=b\,\uu{e}_{1}$}. The dislocation is supposed to be spread uniformly on an arbitrary surface $S_0$ \aj{(of dimensions the layer's height and dislocation's width)}{, with normal $\uu{n}$,} so that the {Burgers vector and the} dislocation density tensor {are related by}
\begin{equation}
\uu{b} = \int_{S_0} \bm{\alpha}.\uu{n}~{\rm d}S.
\end{equation}
For the {2D} problem {the \aj{initial} dislocation density components are}
\begin{equation}
\alpha_{11}^0 = \alpha_{12}^0 = \frac{b}{S_0}.
\end{equation}
\aj{Here the surface $S_0$ is supposed to be square with the layer's height and dislocation's width both taken equal to $10b$. Material data corresponding to aluminum are
considered: {the norm of the Burgers vector is $b=0.286$ nm, the viscous drag coefficient is $\eta= 10^5$
Pa.s.m$^{-1}$ \citep{cho_mobility_2017}}, the elastic constants are $\mu=26.1$ GPa and $\lambda=46.3$ GPa, and the density is $\rho = 2700$ kg.m$^{-3}$. With the considered surface ${S_0}=100b^2$, the dislocation densities are
$\alpha_{11}^0 = \alpha_{12}^0 = 3.5 \times 10^{7}$ m$^{-1}$, (or equivalently $\tilde{\alpha}_{11}^0 = \tilde{\alpha}_{12}^0 = 10^{-2}$). Finally, a uniform remote stress $\tau_{13} = V_s\eta \approx \pm 310$ MPa is considered so that the celerity of dislocations reads $v_0 = \pm 1$.}

\subsection{1D example: annihilation of dislocations}
As a first example, we consider a 1D version of the transport equation:
\begin{equation}\label{}
\frac{\partial {U}_{13}^{\rm p}}{\partial \tilde{t}} + v_0 \left|\frac{\partial {U}_{13}^{\rm p}}{\partial \tilde{x}_1}\right| =0.
\end{equation}
In this case, only \aj{straight parallel edge dislocations are considered in the slip plane, that is $\alpha_{11}=0$ and $\alpha_{12}=\alpha_{12}(x_1)$}. Physical phenomena such as the propagation of a sole dislocation and annihilation of  two dislocations of opposite sign can be investigated using this 1D equation. Here we focus on the process of dislocation annihilation, resulting of the shock of two dislocations of opposite sign (see Figure \ref{fig:1D_Anni}). This prototype equation was investigated in previous works \citep{varadhan_dislocation_2006,djaka_numerical_2015,xia_elazab_2015} and was found to generate numerical \aj{artifacts} such as oscillations and damping.

\begin{figure}[!ht]
\centering
\subfloat[]{\includegraphics[width=7.5cm]{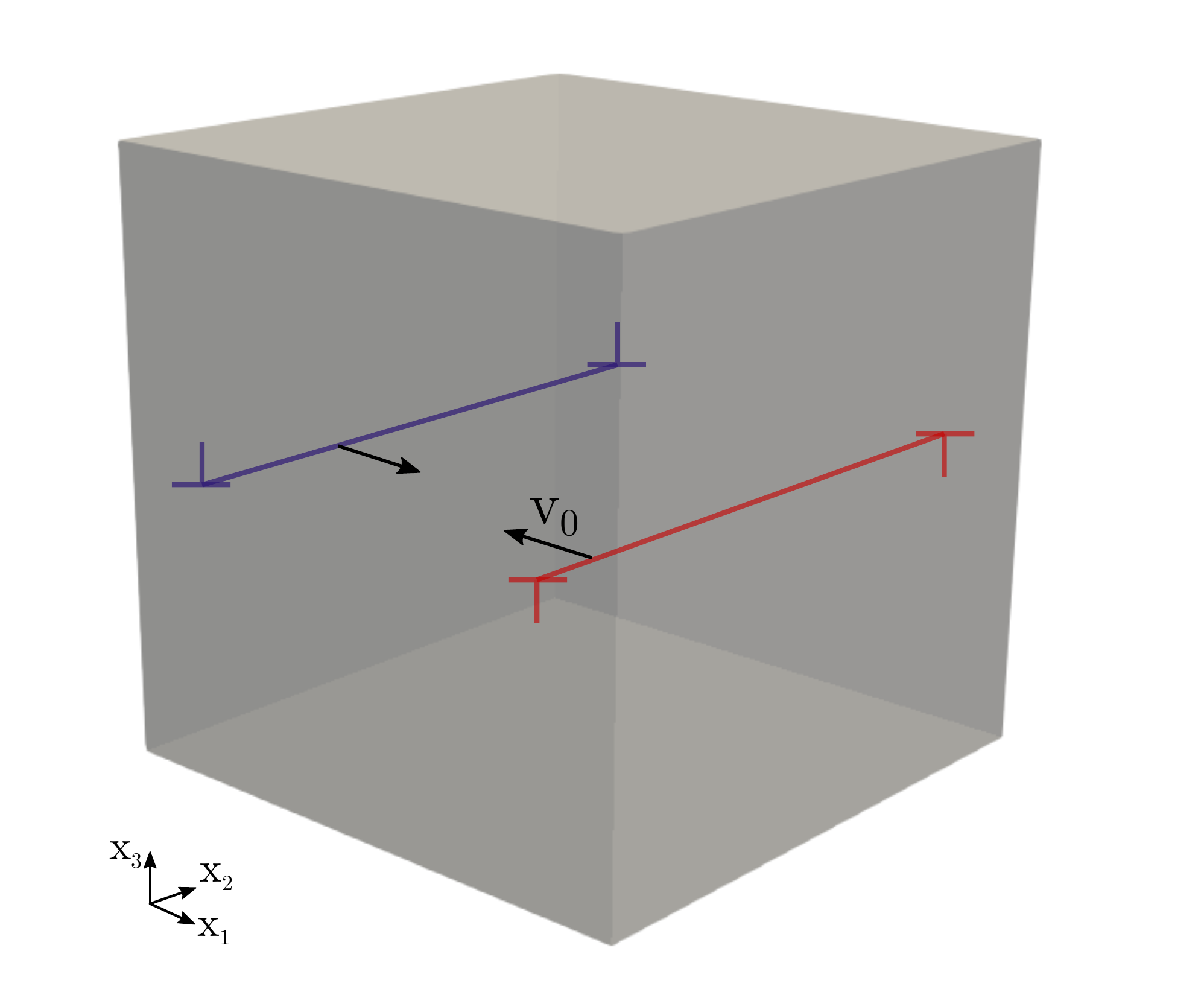}}\hspace{0.5cm}
\subfloat[]{\includegraphics[width=7.5cm]{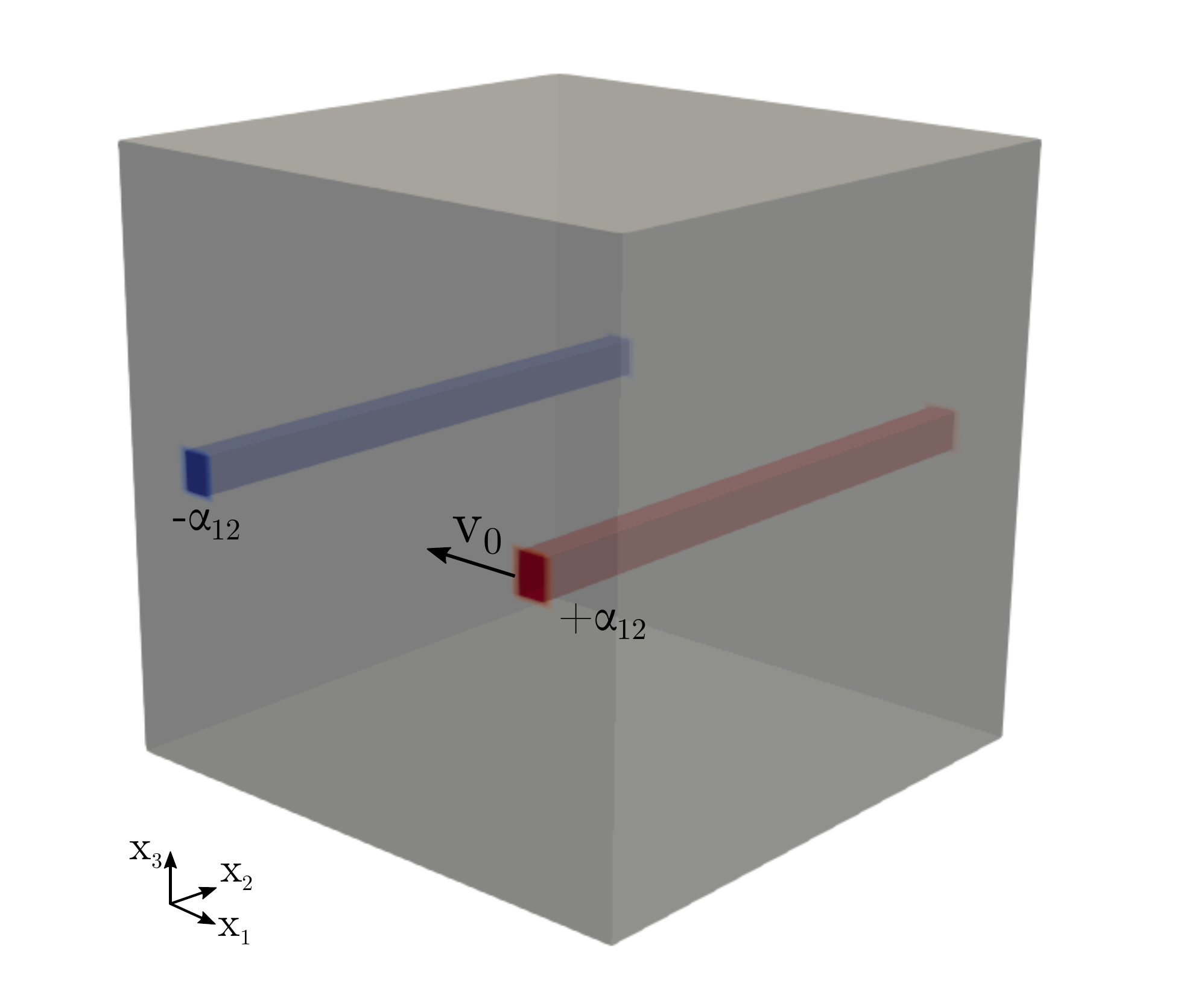}}
\caption{Mechanism of annihilation of dislocations. (a) Discrete edge dislocations, (b) 3D density of edge dislocations.}
\label{fig:1D_Anni}
\end{figure}

\aj{Two initial edge dislocation density distributions, modeled by half-square waves of amplitude $ \tilde{\alpha}_{12}^0 = \pm 10^{-2}$ are embedded in a uniform velocity field $v_0 = -1$. The unit-cell of size $320b$ is discretized on a regular grid of 2048 pixels, so the spatial scale is $\Delta {x}_1 \approx 0.16b$, or equivalently $\Delta \tilde{x}_1 \approx 0.16$. In this case, the CFL number considered is
\begin{equation}
|v_0| \frac{\Delta \tilde{t}}{\Delta \tilde{x}_1} = 0.25
\end{equation}
so the dimensionless time step is $\Delta \tilde{t} \approx 0.04$, or equivalently $\Delta {t} \approx 3.6\times10^{-15}$ s.} \\

The results are represented in Figure \ref{fig:1D_Anni_FDM} at several time steps. The two half-square waves move towards each other and {collide} when they meet at the center of the unit-cell. In this {unidimensional} case, the evolution of the dislocation density predicted by the numerical scheme coincides almost exactly with the exact evolution calculated using the method of characteristics. (The exact solution is not represented since it would be indistinguishable from the numerical solution). In particular the dislocation densities are transported without damping and oscillation. This is in contrast with previous works where strong oscillations and damping {were} observed, requiring numerical heuristic methods such as diffusion terms \citep{varadhan_dislocation_2006} or spectral filters \citep{djaka_numerical_2015}{, even in the case of smooth sinusoidal signals.}

\begin{figure}[!ht]
\centering
\subfloat[]{\includegraphics[height=3.8cm]{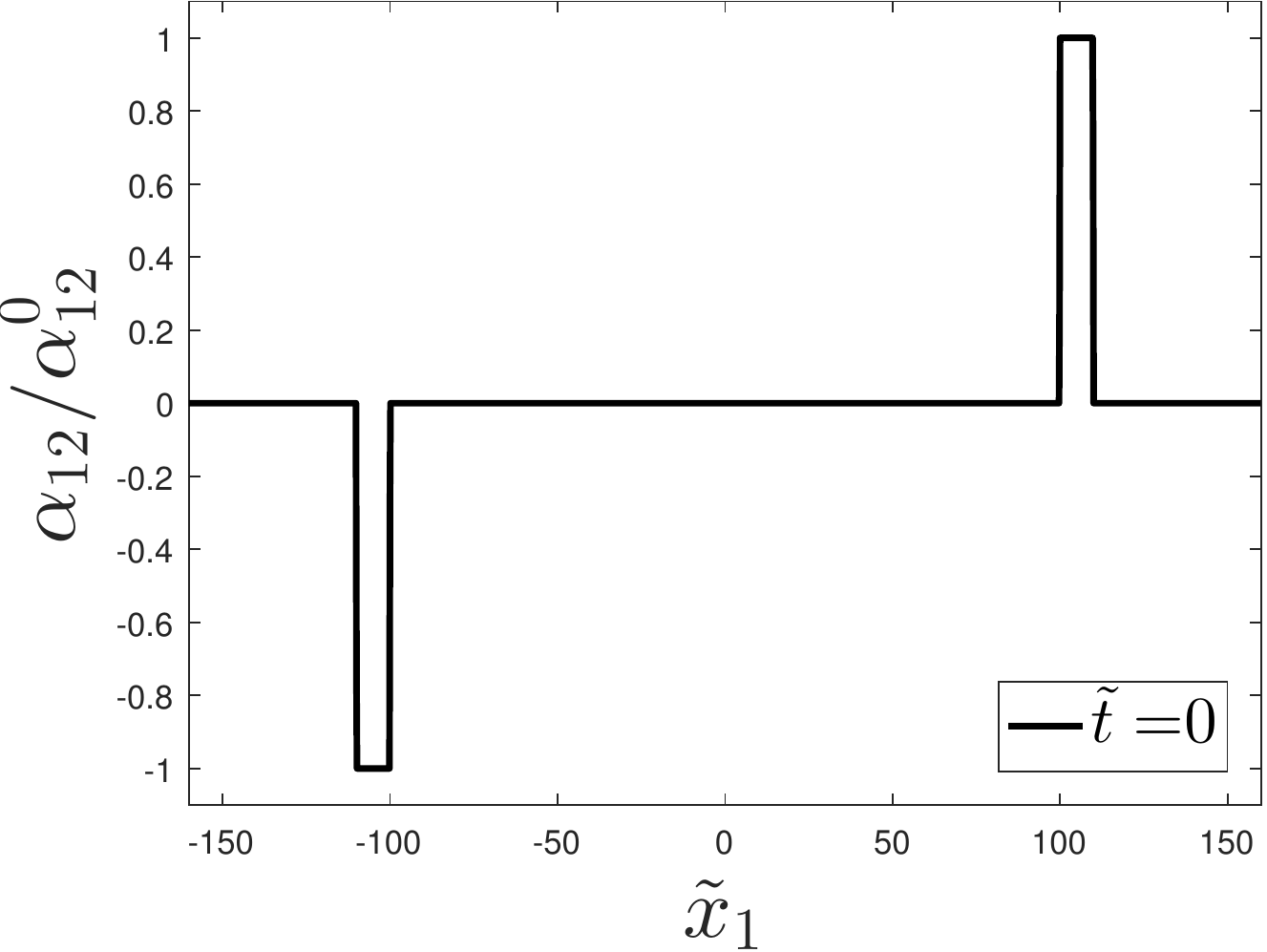}}\hspace{0.1cm}
\subfloat[]{\includegraphics[height=3.8cm]{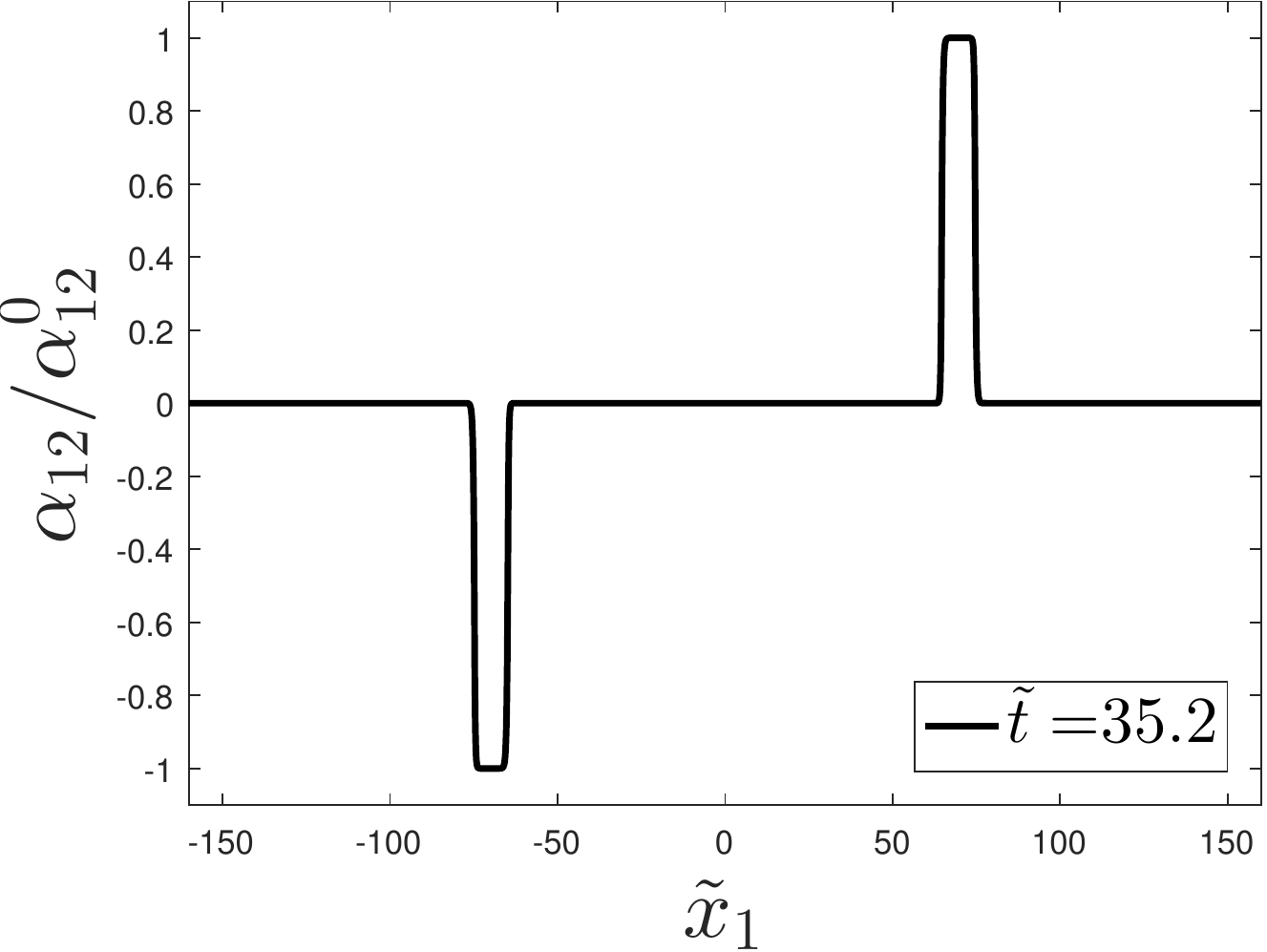}} \hspace{0.1cm}
\subfloat[]{\includegraphics[height=3.8cm]{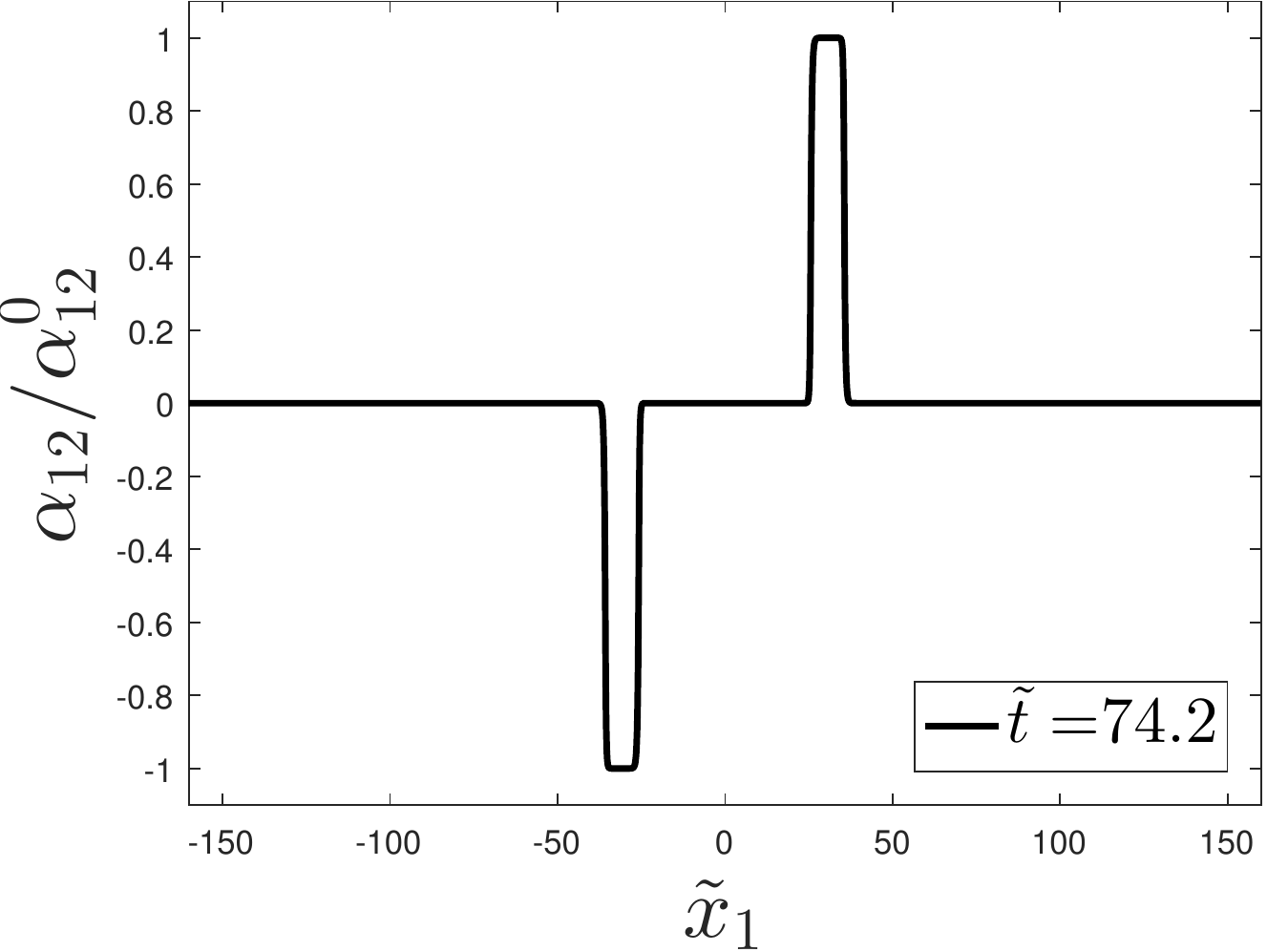}} \\
\subfloat[]{\includegraphics[height=3.8cm]{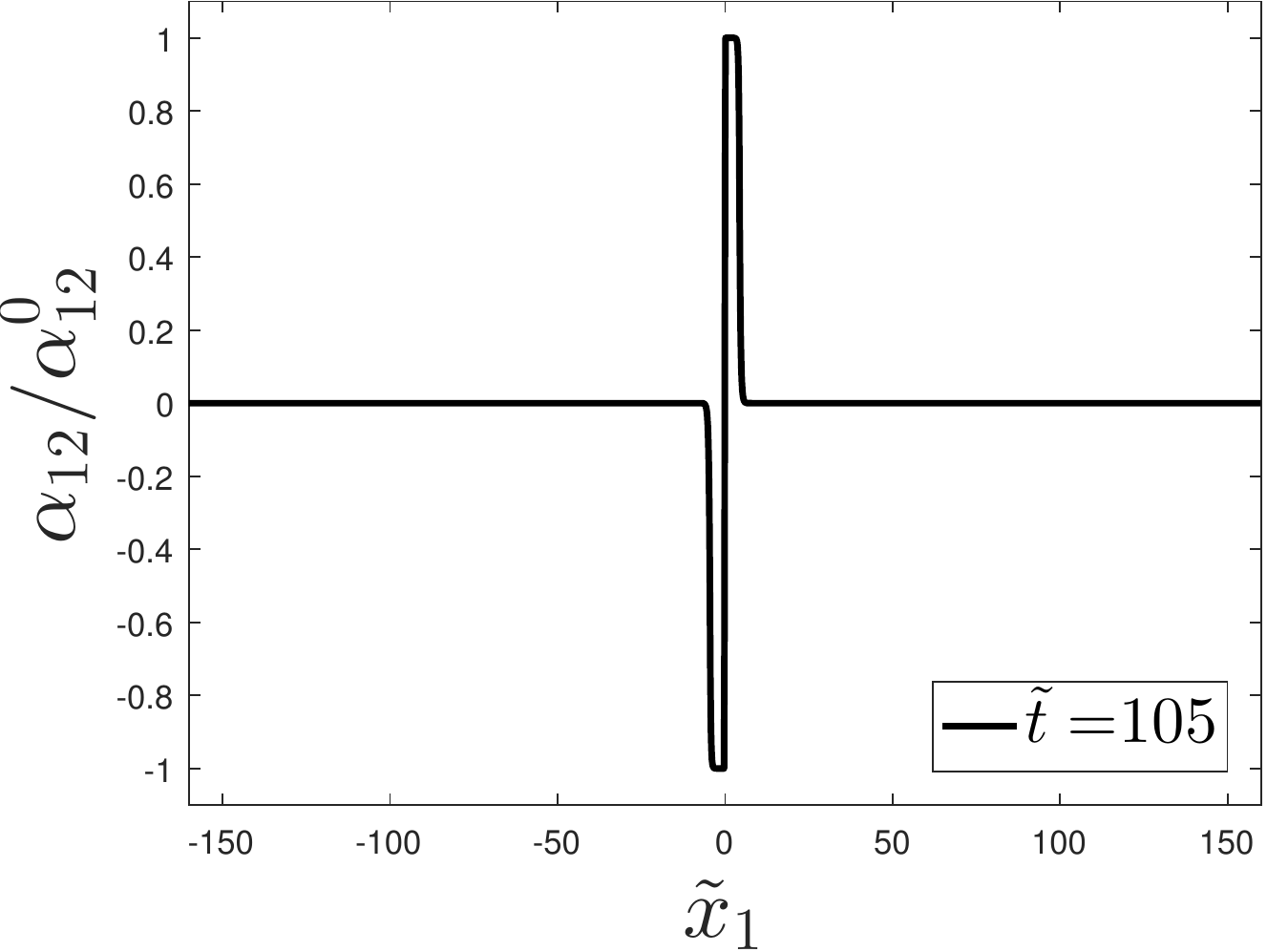}} \hspace{0.1cm}
\subfloat[]{\includegraphics[height=3.8cm]{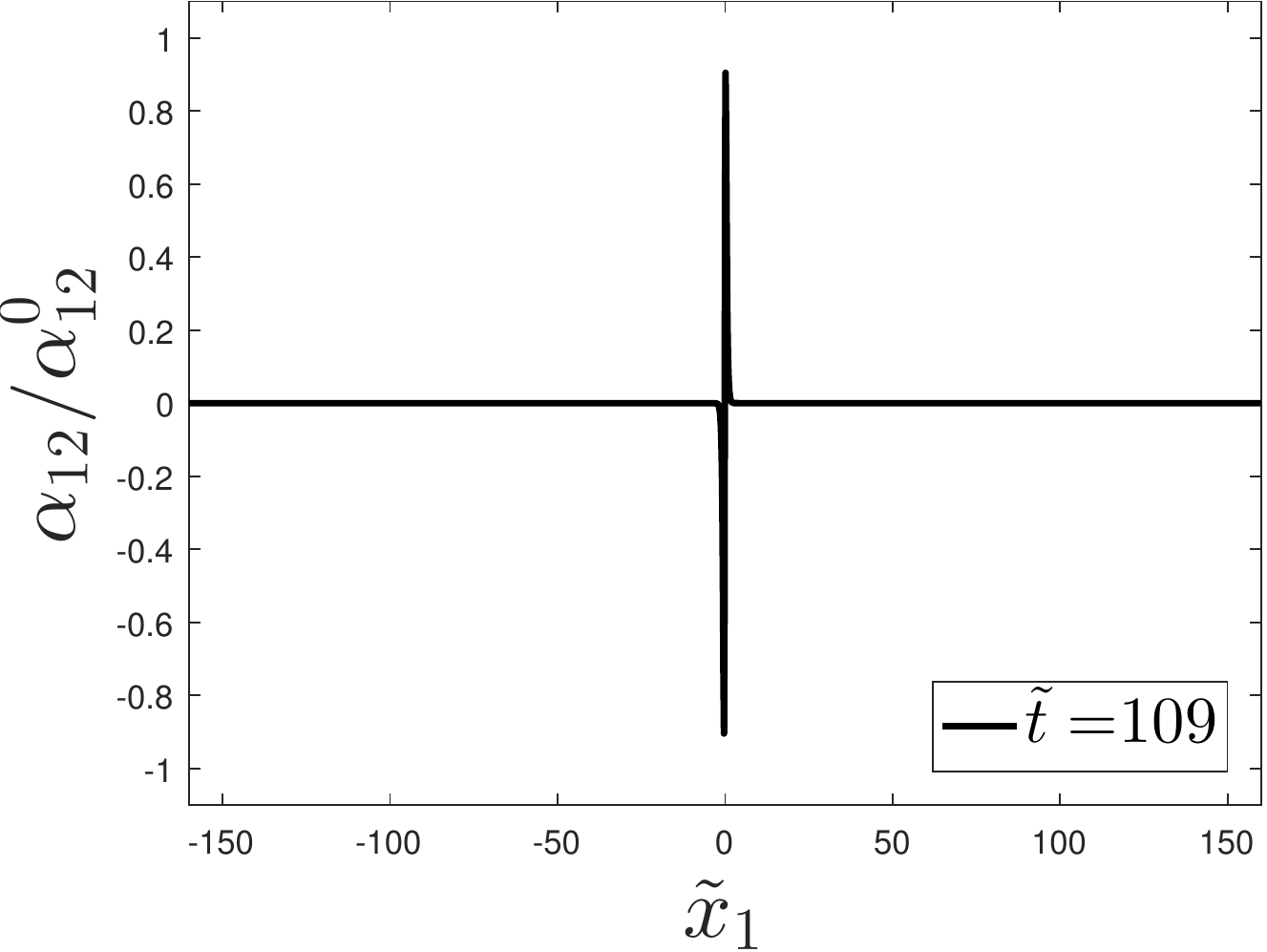}}\hspace{0.1cm}
\subfloat[]{\includegraphics[height=3.8cm]{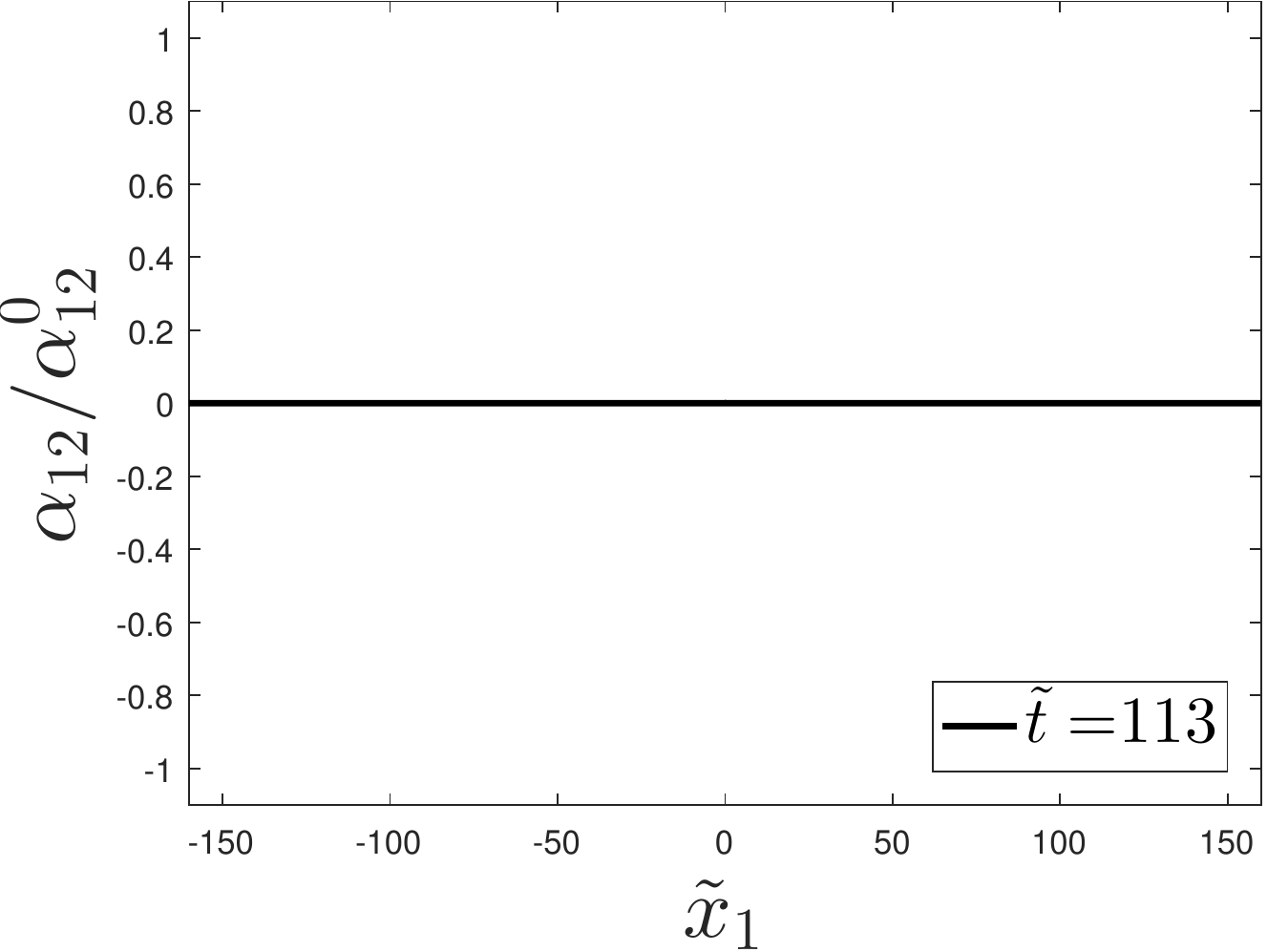}}
\caption{Evolution of the edge dislocation density distribution $\alpha_{12}/\|\alpha_{12}^0\|_{\rm max}$ in the process of annihilation of dislocations. (a) $\tilde{t}=0$, (b) $\tilde{t}=35.2$, (c) $\tilde{t}=74.2$,  (d) $\tilde{t}=105$ (e) $\tilde{t}=109$ (f) $\tilde{t}=113$.}
\label{fig:1D_Anni_FDM}
\end{figure}

\subsection{\aj{2D examples: expansion dislocation loops}}
As a second example, we consider the 2D version of the transport equation:
\begin{equation}\label{}
\frac{\partial {U}_{13}^{\rm p}}{\partial \tilde{t}} + v_0 \sqrt{  \left( \frac{\partial {U}_{13} }{\partial \tilde{x}_1} \right)^2 + \left( \frac{\partial {U}_{13} }{\partial \tilde{x}_2} \right)^2 }=0.
\end{equation}
In this case, both edge and screw dislocations {($\alpha_{12}\ne 0, \alpha_{11}\ne 0$)} are considered in the slip plane. Physical phenomena such as the expansion \aj{and shrinkage} of planar dislocation loops can be investigated with this equation. 

\subsubsection{\aj{Smooth circular dislocation loop}}\label{sec:smooth_loop}
Here we focus on the process of expansion of a \aj{smooth} circular dislocation loop (see Figure \ref{fig:2D_Loop}).
This example was {also} investigated in previous works where important spreading and damping was observed.

\begin{figure}[!ht]
\centering
\subfloat[]{\includegraphics[width=6.5cm]{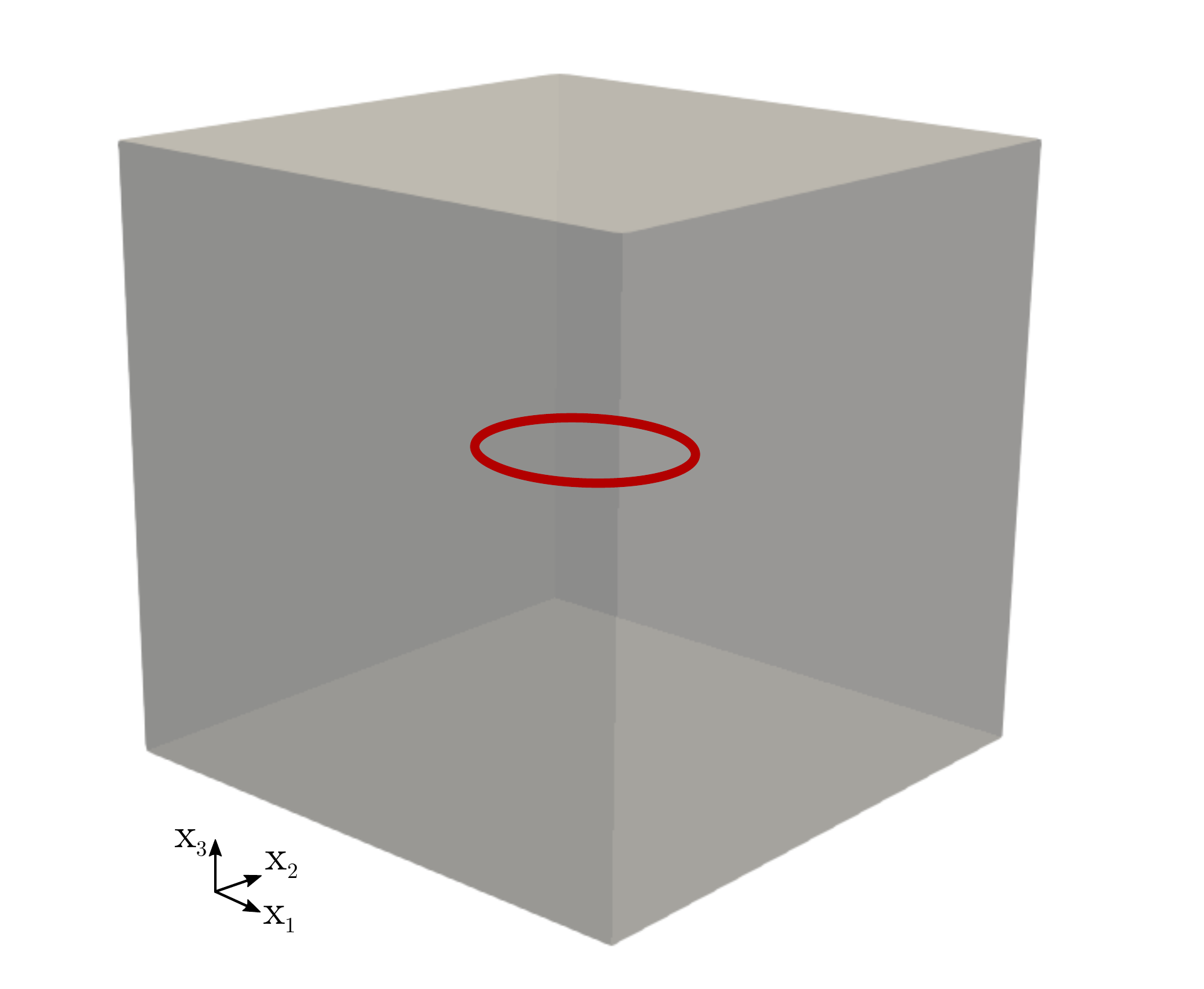}}\hspace{0.5cm}
\subfloat[]{\includegraphics[width=6.5cm]{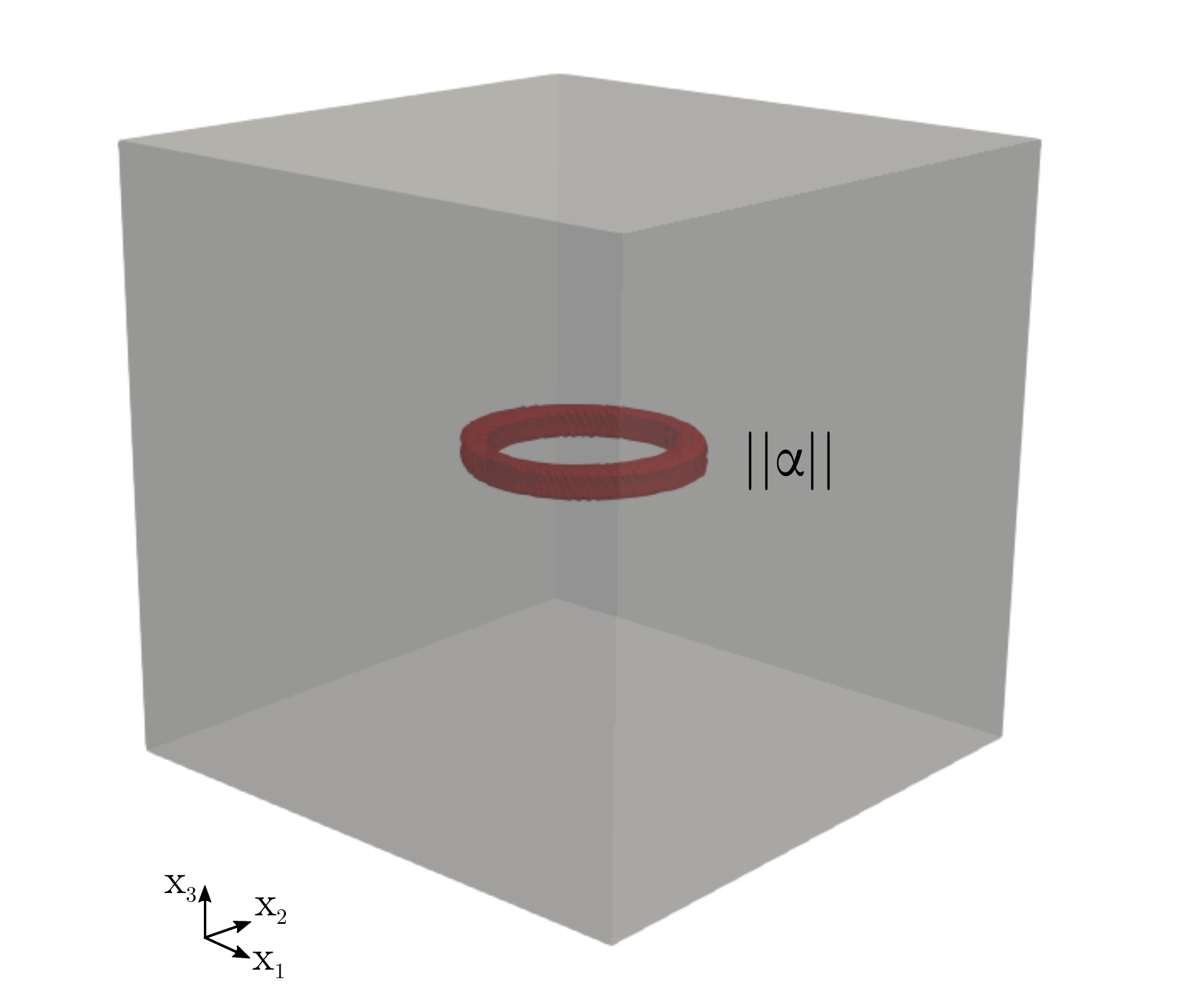}}\hspace{0.5cm} \\[0.5cm]
\subfloat[]{\includegraphics[width=11.5cm]{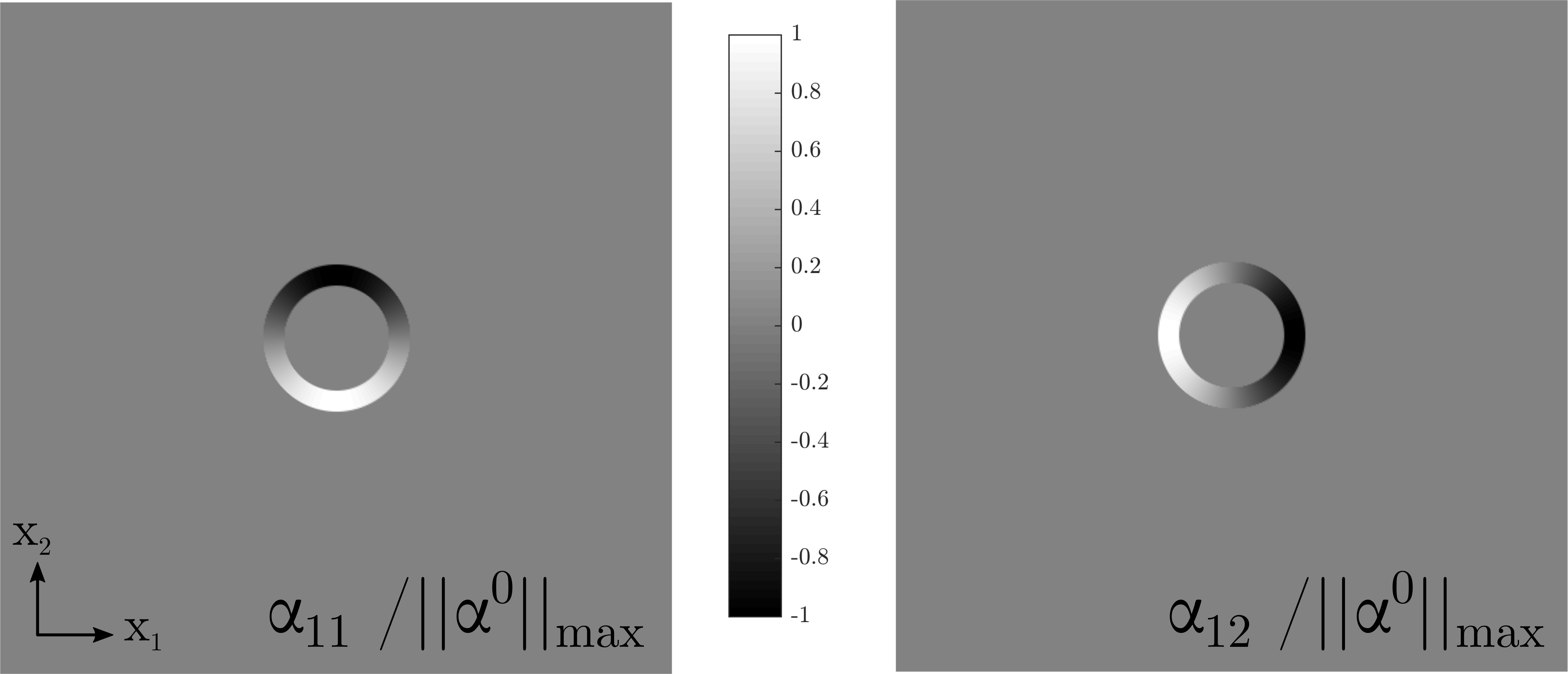}}
\caption{Mechanism of expansion of a circular dislocation loop. (a) Discrete dislocation line, (b) 3D density of dislocation, (c) 2D problem considered.}
\label{fig:2D_Loop}
\end{figure}

\aj{The initial dislocation density, \aj{characterized by $\|\alpha^0\| = \sqrt{\left(\alpha_{11}^0\right)^2+\left(\alpha_{12}^0\right)^2} = 3.5 \times 10^7$ m$^{-1}$} and represented in Figure \ref{fig:2D_Loop}, is again embedded in a uniform velocity field $v_0 =\aj{-1}$. The unit-cell is discretized on a regular grid of $512 \times 512$ pixels, so the spatial scale is $\Delta x = 0.62b$, (or $\Delta \tilde{x}=0.62$). The CFL number considered is {still}
\begin{equation}
|v_0| \frac{\Delta \tilde{t}}{\Delta \tilde{x}_1} = 0.25
\end{equation}
so the dimensionless time step is $\Delta \tilde{t}\approx0.16$, or equivalently $\Delta t \approx 1.44 \times 10^{-14}$ s.}

The results are represented in Figure \ref{fig:expansion_ronde} at several time steps. The mechanism of expansion is well reproduced by the scheme {and,} again, the dislocation density is transported without any damping and spreading, in contrast with previous works.


\begin{figure}[!ht]
\centering
\includegraphics[width=12cm]{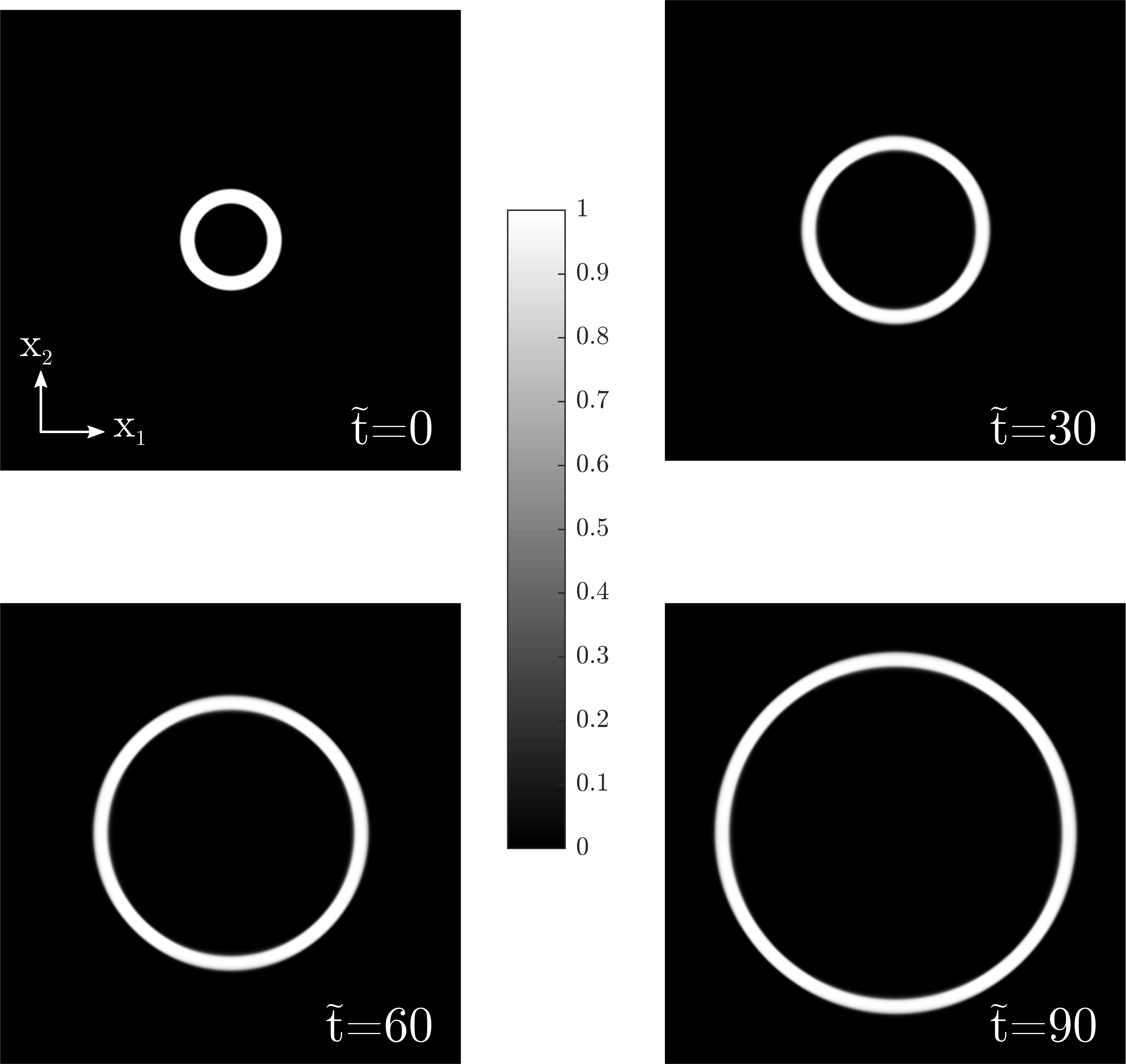}
\caption{\aj{Evolution of the dislocation density $\|{\alpha}\| / \|{\alpha}^0\|_{\rm max}$ in the process of expansion of a smooth circular loop.}}
\label{fig:expansion_ronde}
\end{figure}

\clearpage
\subsubsection{Polygonal dislocation loop}
\aj{We continue with the case of a dislocation loop with corners as defined in Figure \ref{fig:LoopSquare}. This is an interesting case because it admits non-unique weak solutions \citep{acharya_driving_2003}. In particular, following \cite{varadhan_dislocation_2006}'s comments, some entropy condition needs to be specified in order to choose between the so-called expansion fan solution (a moving corner turns into an arc of constant radius) or the shock solution (a moving corner remains sharp)}. \\

\begin{figure}[!ht]
\centering
\includegraphics[width=11.5cm]{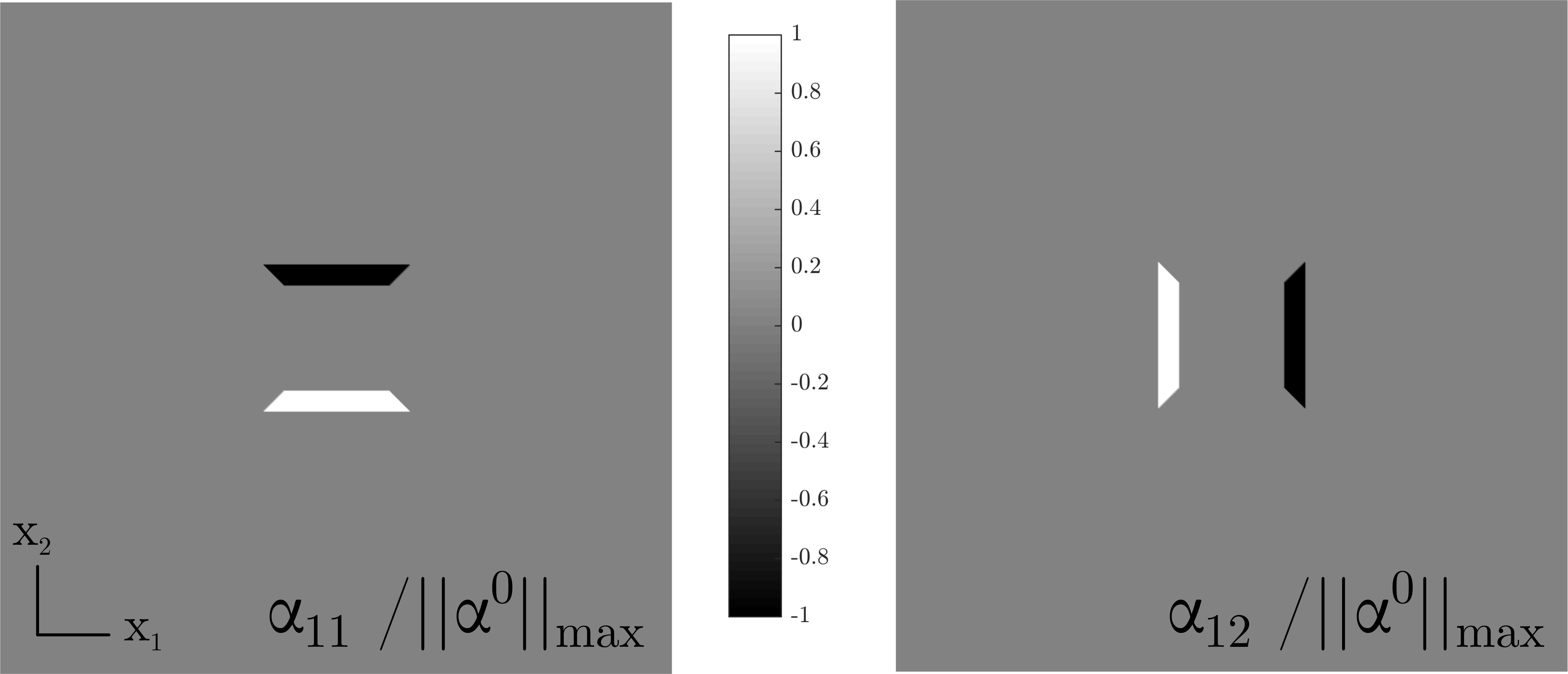}
\caption{2D problem considered in the case of a polygonal dislocation loop.}
\label{fig:LoopSquare}
\end{figure}

\aj{First, we consider the case of expansion of the polygonal loop, which corresponds to a uniform velocity $v_0=-1$. The parameters considered in Section \ref{sec:smooth_loop} are again used. The mechanism of expansion is well reproduced in Figure \ref{fig:expansion_poly} by the scheme without notable damping and spreading. The corners do not stay sharp which means that the scheme automatically chooses the expansion fan solution. It should be noted that if the process is reversed at the end of the expansion by imposing $v_0=-1$, the dislocation loop takes its initial polygonal shape.} \\

\begin{figure}[!ht]
\centering
\includegraphics[width=12cm]{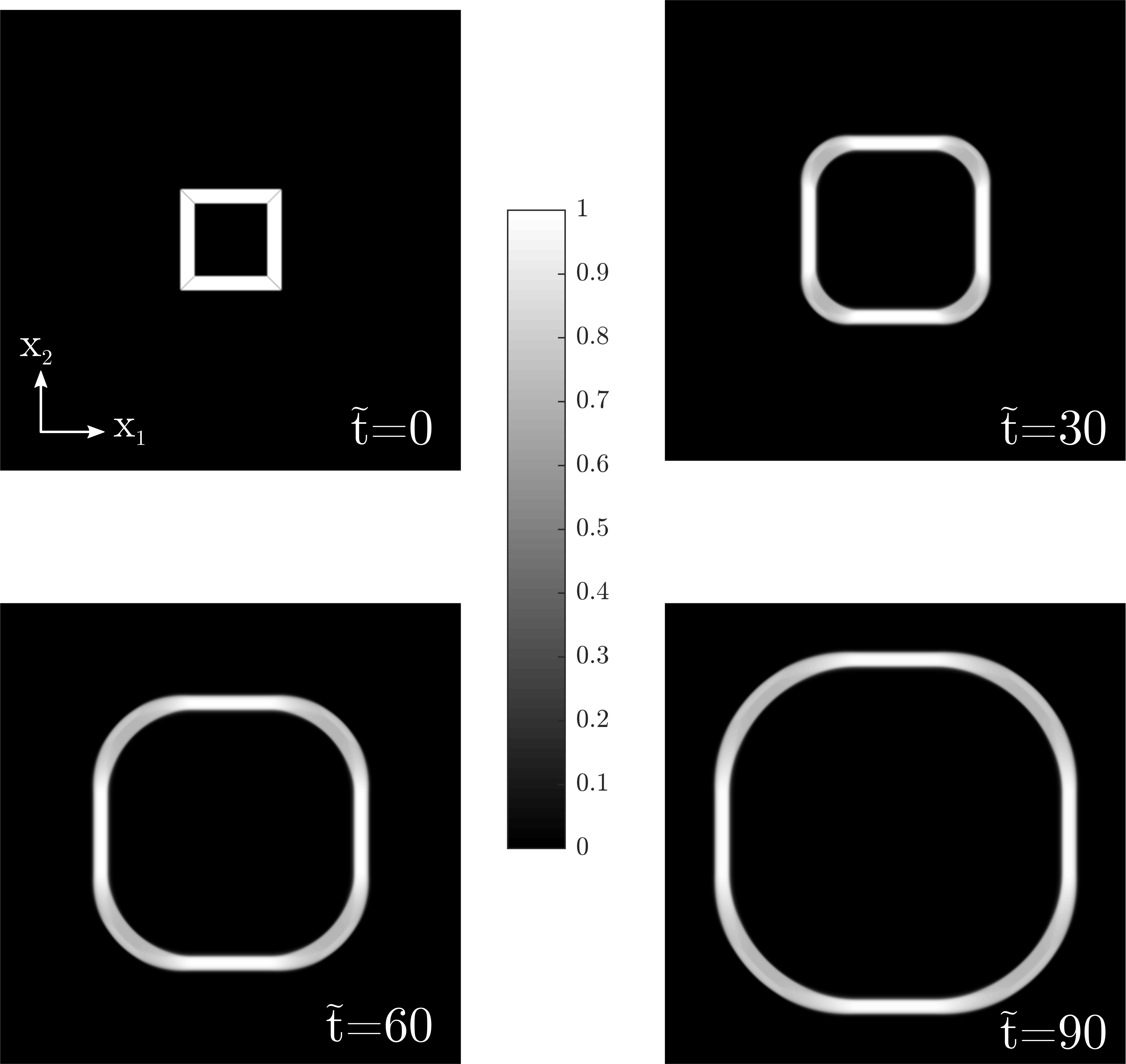}
\caption{\aj{Evolution of the dislocation density $\|{\alpha}\| / \|{\alpha}^0\|_{\rm max}$ in the process of expansion of a polygonal loop.}}
\label{fig:expansion_poly}
\end{figure}

\aj{Then, we consider the case of shrinkage of an initially polygonal loop (defined in Figure \ref{fig:LoopSquare}), which corresponds to a uniform velocity $v_0=1$. (Again, the same other parameters are considered). The mechanism of shrinkage is well reproduced in Figure \ref{fig:contract_poly} by the scheme without regularizing the corners as in the expansion process.}

\begin{figure}[!ht]
\centering
\includegraphics[width=12cm]{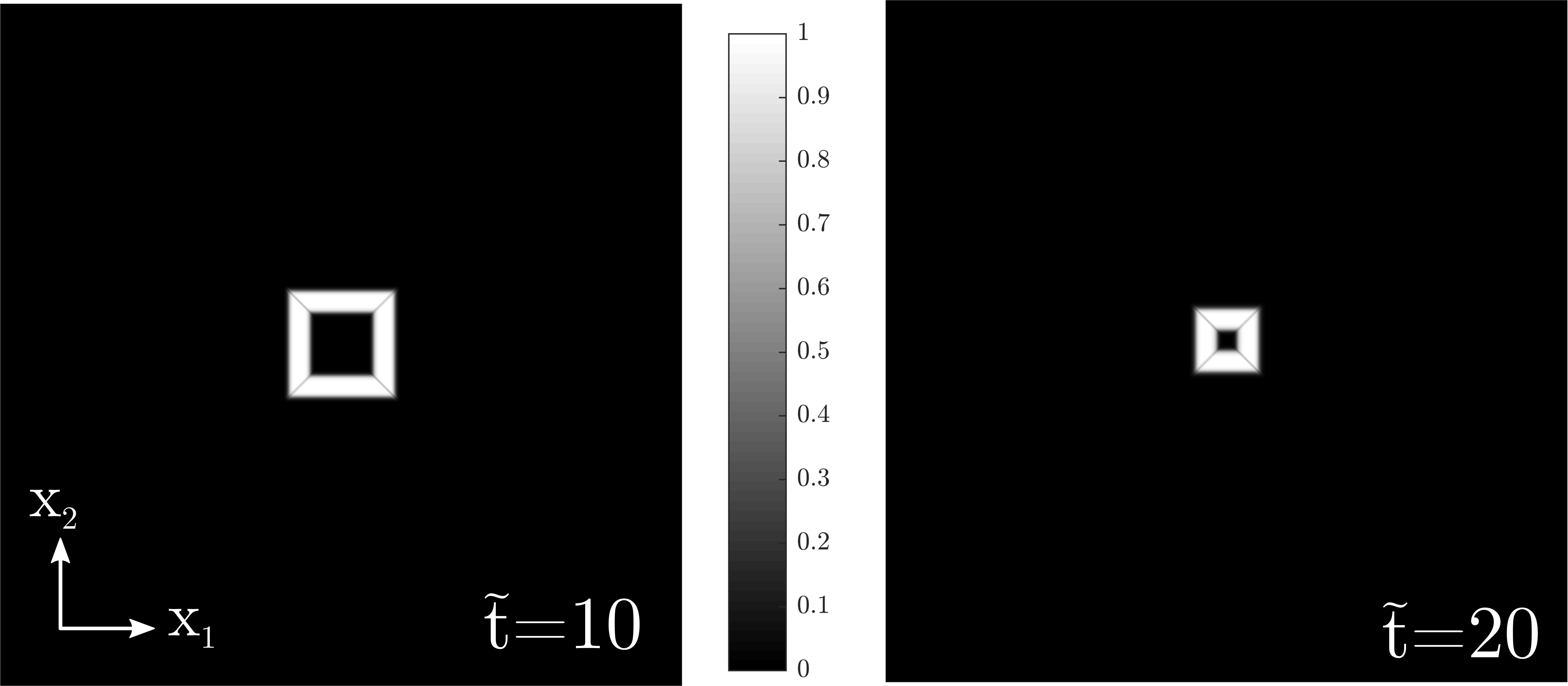}
\caption{\aj{Evolution of the dislocation density $\|{\alpha}\| / \|{\alpha}^0\|_{\rm max}$ in the process of shrinkage of a polygonal loop.}}
\label{fig:contract_poly}
\end{figure}

\clearpage
\subsection{\aj{Discussion}}\label{sec:4.4}
\aj{As shown in Section \ref{sec:4.1}, the transport of $\|{\alpha}\|$ modeled by equation \eqref{eq:HJ2D} is {\it conservative} in the velocity field considered}, which means that no damping and no spreading of $\| \alpha \|$ should be observed. \aj{The numerical scheme considered in this work has permitted to respect this property. The present results {thus} improve the prediction of dislocation motion formulated in {the} field dislocation mechanics framework in a simplified case where the plastic distortion tensor reduces to one component}; indeed  previous works  suffer, \aj{in the same simplified problem}, from numerical discrepancies such as oscillations, spreading and damping of dislocation densities.  {It is pointed out that} predicting correctly the dislocation density motion is of the highest importance in {\it coupled} problems where the magnitude of the dislocation density will determine the stress level. In particular, an inaccurate transport of the dislocation density with damping, spreading and oscillations would induce spurious errors in the predictions of the stress and thus a poor prediction of the elastoplastic mechanical behavior.

\clearpage
\section{Numerical results: coupled problems}\label{sec:results_coupled}

\subsection{Preliminaries}
\aj{The aim of this section is to study numerically the evolution of the pointwise dislocation density tensor by considering the simplified FDM layer problem defined in Section \ref{eq:simplifedlayer} with no a priori assumptions on the velocity of dislocations $v_0$: the stress field is neither constant nor uniform and the stress threshold is strictly positive; the evolution equation \eqref{eq:HJ2D_reso} is coupled to the static problem (relations \eqref{eq:staticdiscr} and \eqref{eq:EqFDMEvolution}).}

\aj{In the sequel, we consider several 3D microstructures since the static problem is solved on a 3D cell made of elastic regions and a layer governed by FDM equations; in all cases, the unit-cell of $320b\times320b\times320b$ is discretized on a regular grid of $256\times 256\times 256$ pixels, so the spatial scale is $\Delta x = 3.58 \times 10^{-1}$ nm. The thickness $h$ of the layer is chosen to be very small ($h=5b$) so that the layer may be seen as a slip plane. This permits to reduce the possible fluctuation of the stress  $\sigma_{13}$ in the $x_3$-direction so that that the average stress $\tau_{13}$ is very close to the stress $\sigma_{13}$. Again, material data corresponding to aluminum are considered (see Section \ref{sec:4.2}). The parameters for the non-convex energy function are taken as follows: $\beta=10^{-8}$ and $\tau_y=1$ MPa. The value of the threshold $\tau_y$ has been taken to coincide with the Peirls stress of aluminum \citep{kamimura_experimental_2013}.}

\aj{Dislocation microstructures (such as a dislocation dipole for instance) produce initially an internal stress field \citep{brenner_numerical_2014}. Thus, an initial microstructure can evolve without any macroscopic stress field applied, due to the local stress field produced by the dislocation density. Consequently we first need to study the possible equilibrium aspects of initial dislocation densities before any mechanical macroscopic loading. It should be noted that, in absence of lattice friction effects, initial microstructures will evolve and the dislocation will spread \citep{zhang_single_2015}; thus equilibrium positions of dislocation field may be allowed only by the introduction of nonconvex energy density functions \citep{zhang_single_2015}. In order to investigate the possible equilibrium of the initial dislocation density field, we study its evolution while keeping a macroscopic zero stress field. Thus, the cell is subjected to a macroscopic loading path ${\bar{\varepsilon}}_{13}$ given by
\begin{equation}
{\bar{\varepsilon}}_{13} = \ds \frac{\moy{U^{\rm p}_{13}}}{2},
\end{equation}
which imposes that the stress ${\bar{\sigma}}_{13}$ is nil. Then, when the microstructure stops evolving, an equilibrium position is reached.}

\aj{Equilibrated microstructures are then subjected to a mechanical loading in order to investigate the evolution of the dislocation density field. An increasing macroscopic strain ${\bar{\varepsilon}}_{13} = \dot{{\bar{\varepsilon}}}_{13} t$ is applied. The strain rate $\dot{{\bar{\varepsilon}}}_{13}$ is chosen low enough so the evolution may be considered as rate-independent.}

\subsection{Evolution of a dislocation loop}
\aj{We consider the previous case of a circular dislocation loop. The initial dislocation density is {characterized by $\|\alpha^0\| = \sqrt{\left(\alpha_{11}^0\right)^2+\left(\alpha_{12}^0\right)^2} = 10^3$ m$^{-1}$} and is represented in Figure \ref{fig:2D_Loop}. The CFL number considered is
\begin{equation}
|v_0| \frac{\Delta \tilde{t}}{\Delta \tilde{x}_1} = 0.25
\end{equation}
where the celerity of dislocation $v_0 =  ({ {\rm cos}\left(  \frac{U^{\rm p}_{13}}{\beta}\right) \tilde{\tau}_y-\tilde{\tau}_{13}})/{\tilde{\eta}} $ depends on the stress level. The time step $\Delta t$ is adjusted to ensure the value of the CFL number.} \\

\aj{First we let the initial microstructure evolve in order to study its possible equilibrium position. The simulation reveals that the dislocation density field slightly oscillates around an equilibrium position that is represented in Figure \ref{fig:rond_equilibre}. It is interesting to note that the dislocation field is somehow ``noisy'' after this equilibrium process. This is due to the fact that the mobility of dislocations oscillates very rapidly in space around the value zero, due to the non-convex energy contribution which implies an alternation of positive, negative and zero velocity.}

\begin{figure}[!ht]
\centering
\includegraphics[width=10.5cm]{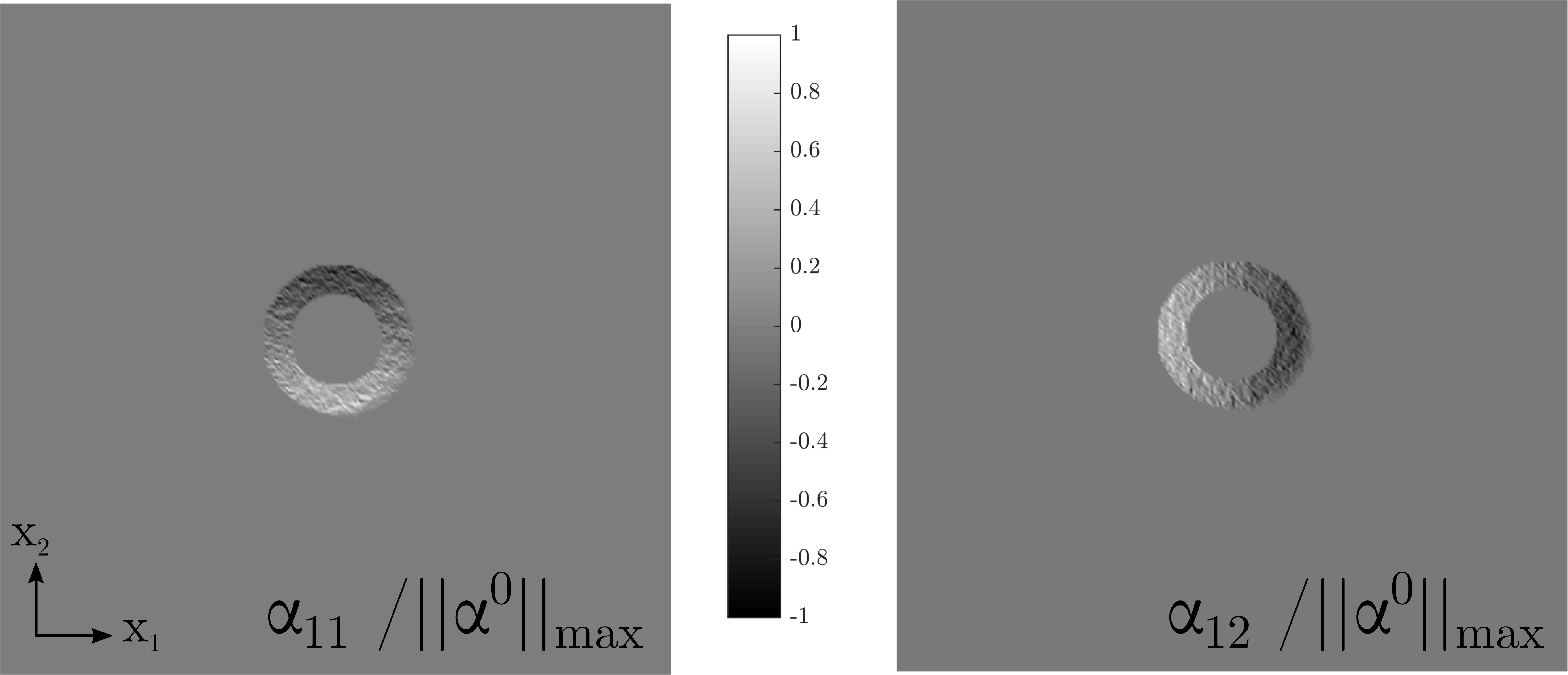}
\caption{Distribution of the dislocation density after equilibrium.}
\label{fig:rond_equilibre}
\end{figure}

\aj{The equilibrated microstructure represented in Figure \ref{fig:rond_equilibre} is then subjected to an increasing macroscopic strain. The results are represented in Figure \ref{fig:expansion_coupled} at several strains: ${\bar{\varepsilon}}_{13} =0$, ${\bar{\varepsilon}}_{13} =5.5\times 10^{-5}$, ${\bar{\varepsilon}}_{13} =8.4\times 10^{-5}$ and ${\bar{\varepsilon}}_{13} =1.1 \times 10^{-4}$. The increase of the total strain ${\bar{\varepsilon}}_{13}$ induces an expansion of the circular loop which is due to an increase of the local stress field. In contrast with the uncoupled problem, the velocity of the dislocation is not prescribed here and is only a consequence of the local mechanical state induced by the macroscopic straining and the dislocation density. It is interesting to note that the noisy effect observed in the equilibrated field disappears during the loading because lattice effects are less dominant when the dislocation starts moving. It is worth noting that the dislocation remains compact during the evolution.}

\begin{figure}[!ht]
\centering
\includegraphics[width=12cm]{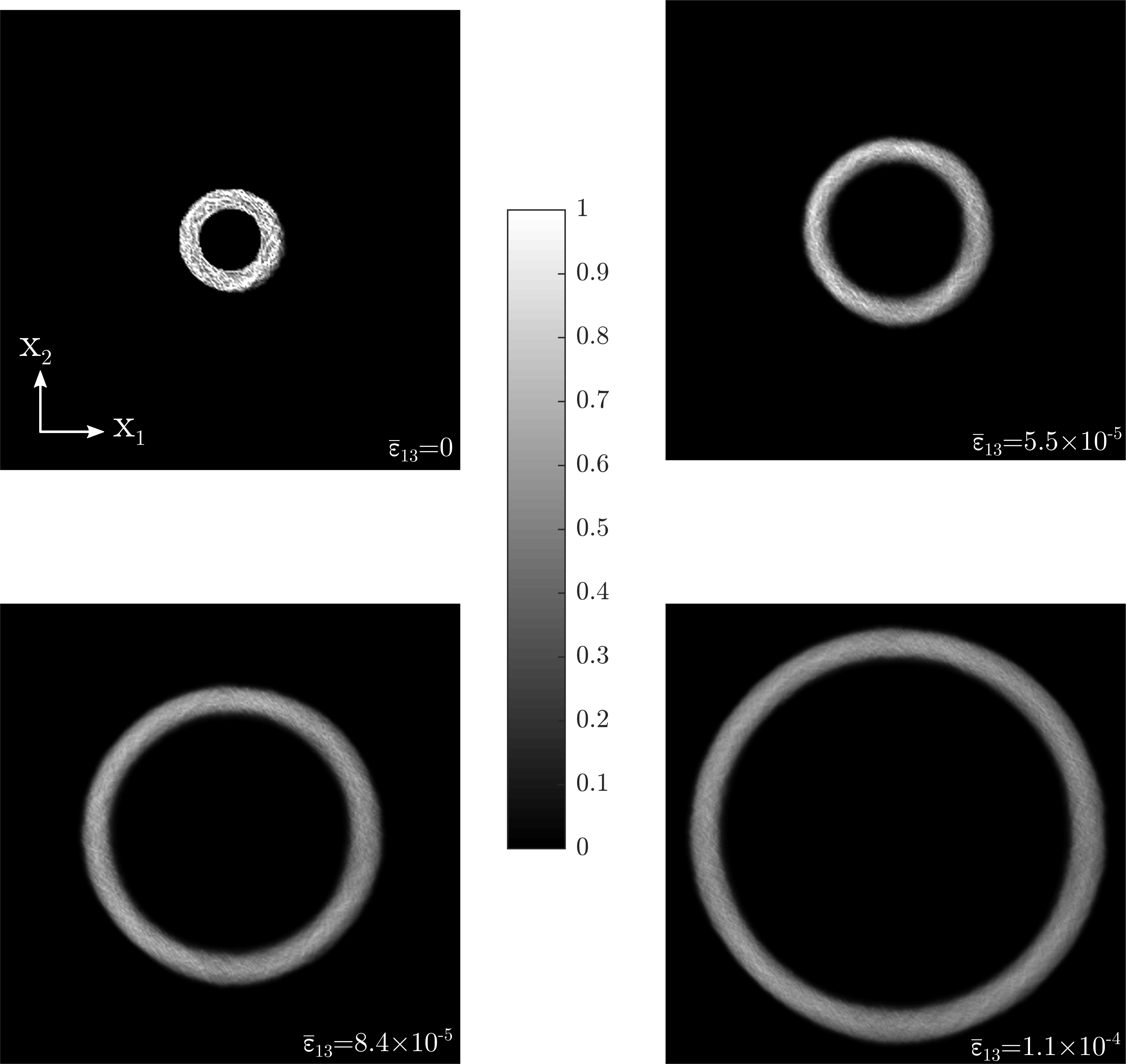}
\caption{\aj{Evolution of the dislocation density $\|{\alpha}\| / \|{\alpha}^0\|_{\rm max}$ of an initially circular loop subjected to an increasing macroscopic strain.}}
\label{fig:expansion_coupled}
\end{figure}

\newpage
\subsection{Orowan's mechanism} 
\aj{As a second example we investigate the interaction between a dislocation loop and a precipitate, which is known as Orowan's mechanism. This mechanism consists in the formation of residual dislocation loops after the bowing of a dislocation around a precipitate. Such mechanism has important consequences on the strength of metallic alloys (i.e. precipitation hardening).  In order to account for the presence of particles, we consider a heterogeneous distribution of the viscous drag coefficient $\eta$. We assume here that particles do not allow the motion of dislocations so they can be modeled by an infinite value of $\eta$ which implies that the dislocation velocity \eqref{eq:defV} is zero.  Two initial distributions are considered, one with dual symmetrical precipitates (see Figure \ref{fig:precipitates}(a)) and the second with a random distribution of precipitates (see Figure  \ref{fig:precipitates}(b)). We consider as before a circular dislocation loop whose equilibrium position is given in Figure \ref{fig:rond_equilibre}.} 

\begin{figure}[!ht]
\centering
\subfloat[]{\includegraphics[width=4.5cm]{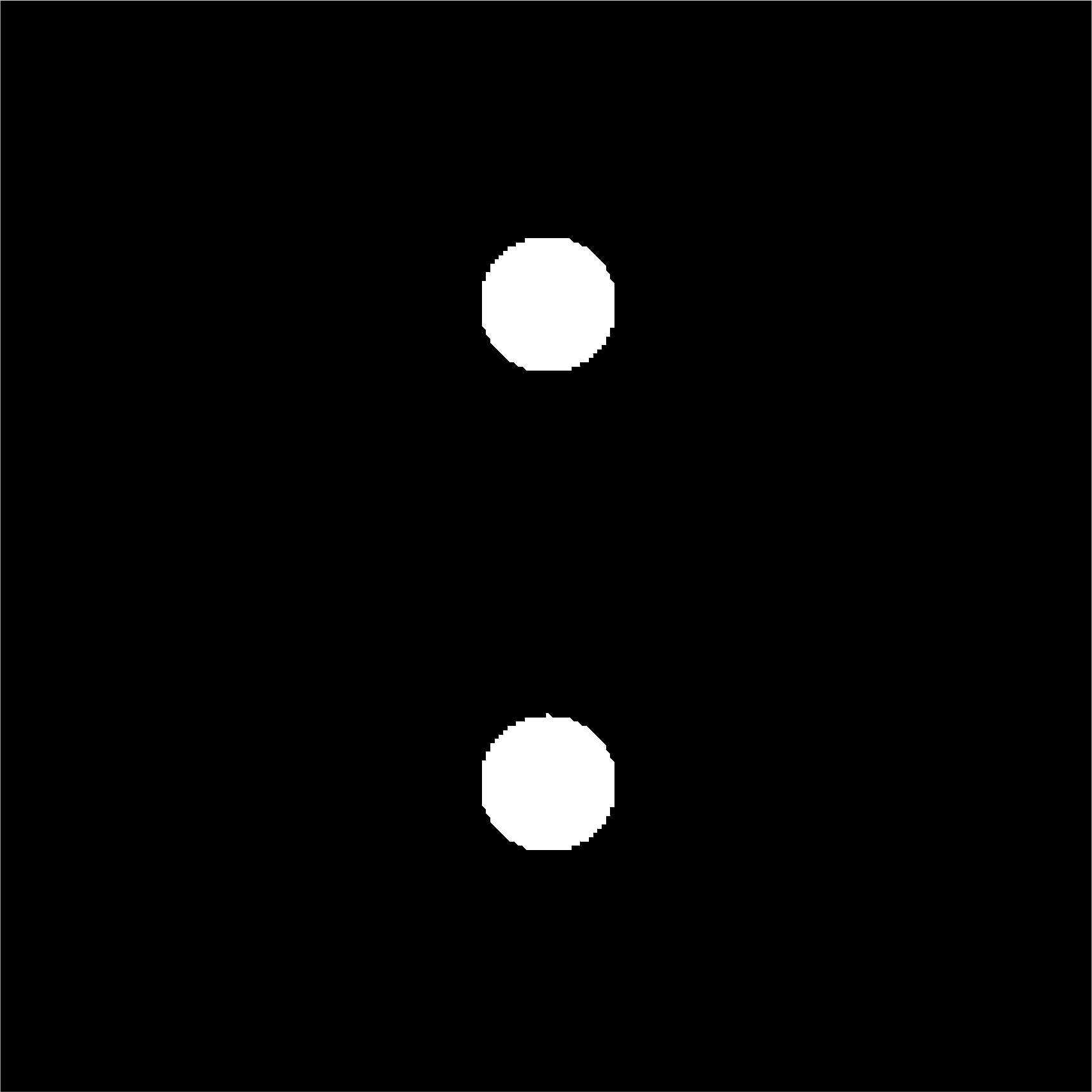}}\hspace{0.5cm}
\subfloat[]{\includegraphics[width=4.5cm]{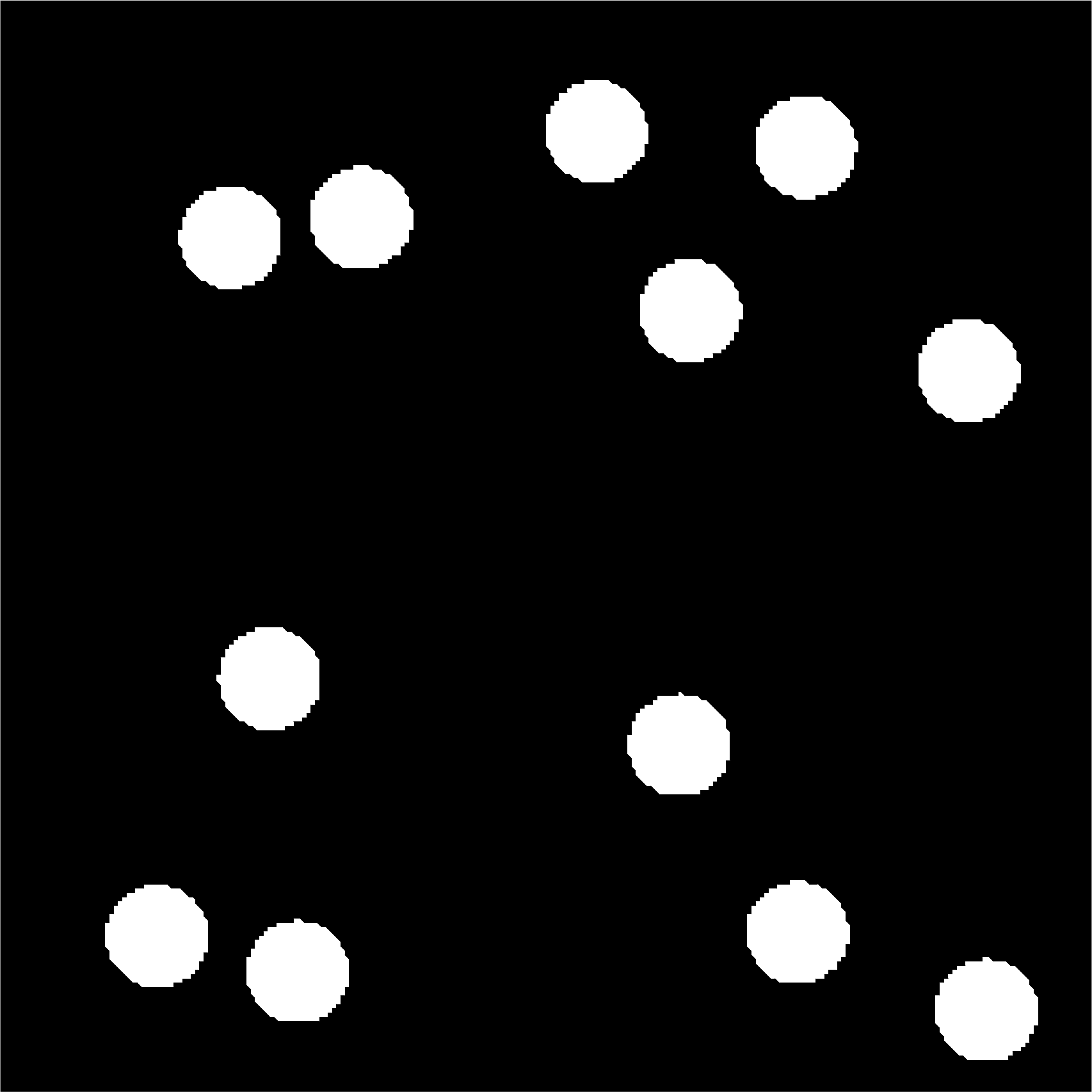}}
\caption{Distribution of the viscous drag coefficient $\eta$ in the case of (a) dual symmetrical precipitates and (b) a random distribution of precipitates. The white color corresponds to $\eta=\infty$ and the black color corresponds to $\eta= 10^5$ Pa.s.m$^{-1}$.}
\label{fig:precipitates}
\end{figure}

\aj{The equilibrated microstructure is then subjected to an increasing macroscopic strain ${\bar{\varepsilon}}_{13}$. In the case of dual symmetrical precipitates, the results are represented in Figure \ref{fig:dualprecip} at several strains: ${\bar{\varepsilon}}_{13} =5.5\times 10^{-5}$, ${\bar{\varepsilon}}_{13} =8.4\times 10^{-5}$, ${\bar{\varepsilon}}_{13} =1.1 \times 10^{-4}$ and ${\bar{\varepsilon}}_{13} =1.2 \times 10^{-4}$. The simulations show that the dislocation loop cuts itself in two parts while it gets around the precipitate. Once the precipitate is passed, the two parts of the dislocation density merge and form again a loop. After this process, a residual dislocation loop remains around the precipitate.}

\begin{figure}[!ht]
\centering
\includegraphics[width=12cm]{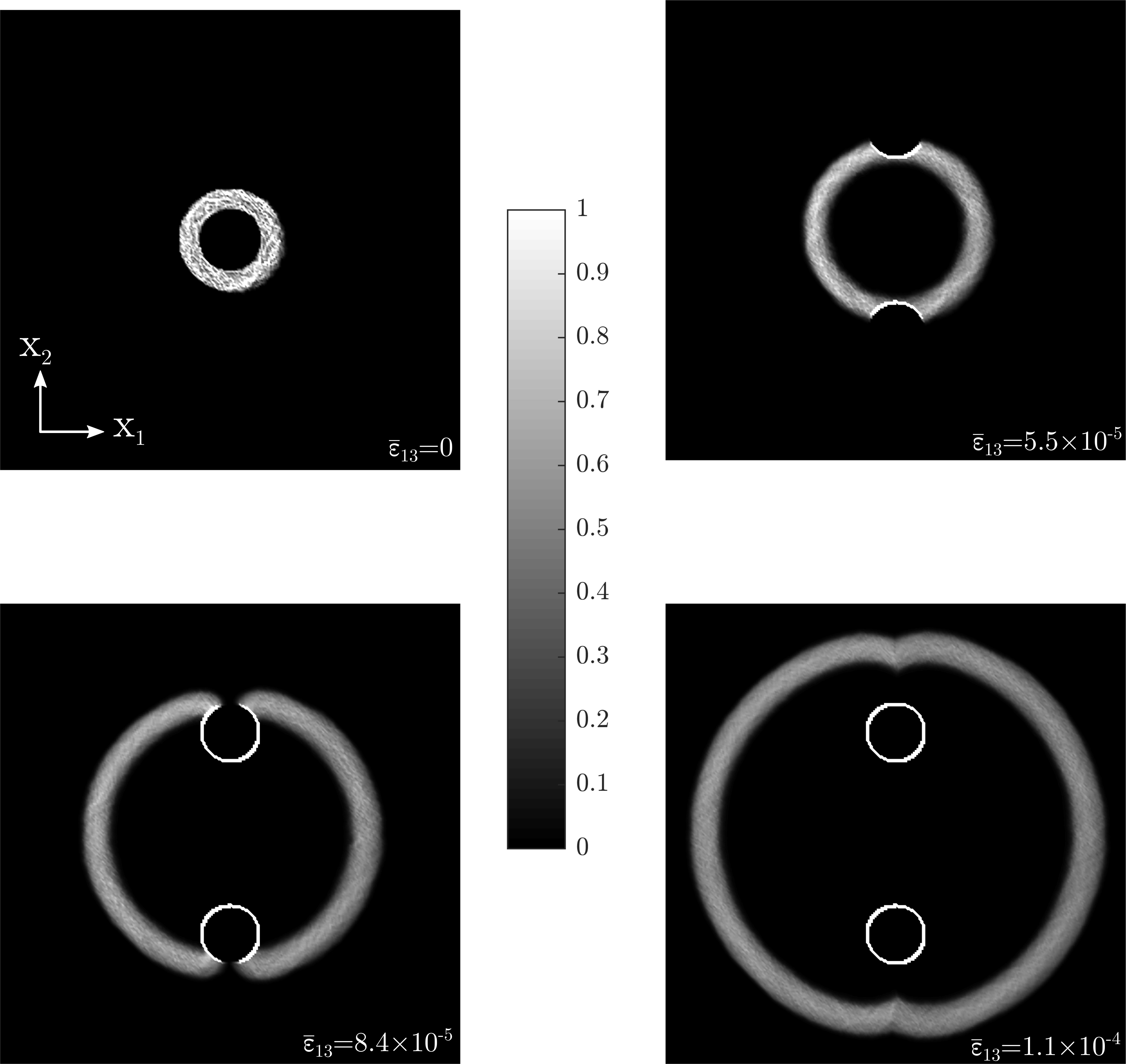}
\caption{\aj{Evolution of the dislocation density $\|{\alpha}\| / \|{\alpha}^0\|_{\rm max}$ of an initially circular loop subjected to an increasing macroscopic strain in the case of dual symmetrical precipitates.}}
\label{fig:dualprecip}
\end{figure}

\aj{In the case of a random distribution of precipitates, the results are represented in Figure \ref{fig:randomprecip} at several strains: ${\bar{\varepsilon}}_{13} =0$ (equilibrium), ${\bar{\varepsilon}}_{13} =5.5\times 10^{-5}$, ${\bar{\varepsilon}}_{13} =8.4\times 10^{-5}$ and ${\bar{\varepsilon}}_{13} =1.1 \times 10^{-4}$. The simulations show that the dislocation loop gets around each precipitate with the same mechanism and a residual dislocation density remains around each precipitate.}

\begin{figure}[!ht]
\centering
\includegraphics[width=12cm]{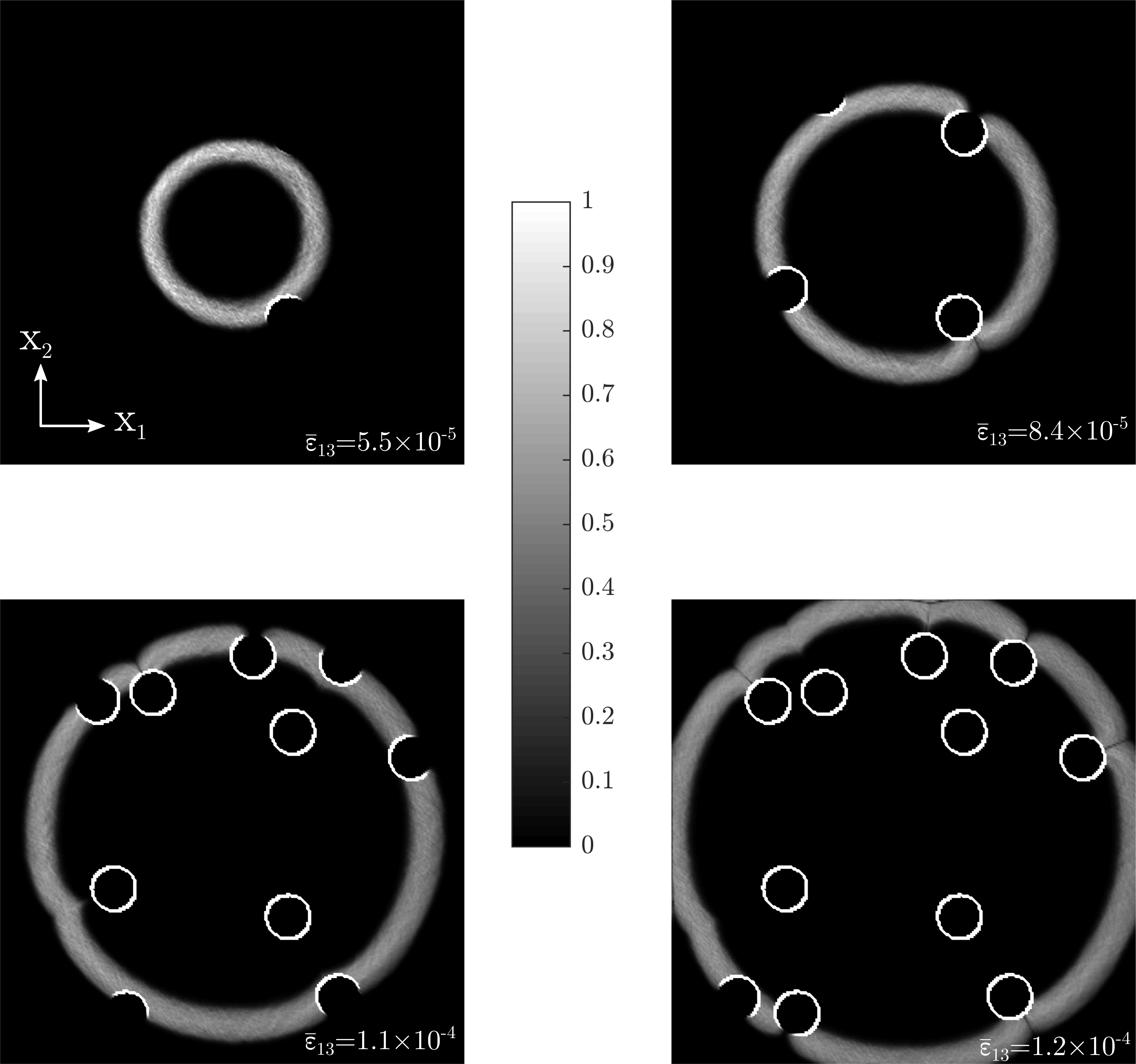}
\caption{\aj{Evolution of the dislocation density $\|{\alpha}\| / \|{\alpha}^0\|_{\rm max}$ of an initially circular loop subjected to an increasing macroscopic strain in the case of a random distribution of precipitates.}}
\label{fig:randomprecip}
\end{figure}

\clearpage
\subsection{Random microstructures} 
\aj{We finally investigate the possible emergence of spatial inhomogeneity of the Nye dislocation field (i.e. dislocation patterning). To do so, we consider the equilibrium and evolution of initially random distributions of the plastic distortion ${U}_{13}^{\rm p}$. A first microstructure (microstructure A) is generated in the plastic layer using a random number generator which ensures that $\|{\alpha}^0\|_{\rm max}=10^3$ m$^{-1}$ and $\overline{{U}}_{13}^{\rm p}=0$ (see Figure \ref{fig:Random_micro}(a)). Since the random number generation is done in each pixel, the distribution of ${U}_{13}^{\rm p}$ is noisy, which may induce damping during the evolution problem. Thus, a second microstructure is generated by applying a smoother \citep{garcia_robust_2010} which permits to keep the same properties ($\|{\alpha}^0\|_{\rm max}=10^3$ m$^{-1}$ and $\overline{{U}}_{13}^{\rm p}=0$) while gaining in smoothness (see Figure \ref{fig:Random_micro}(b)). The aim of this study is only to illustrate the possible emergence of patterning, so no attempt is done here to characterize thoroughly these microstructures in terms of morphology and representativity \citep{jeulin2012}.} \\

\begin{figure}[!ht]
\centering
\subfloat[]{\includegraphics[height=5.cm]{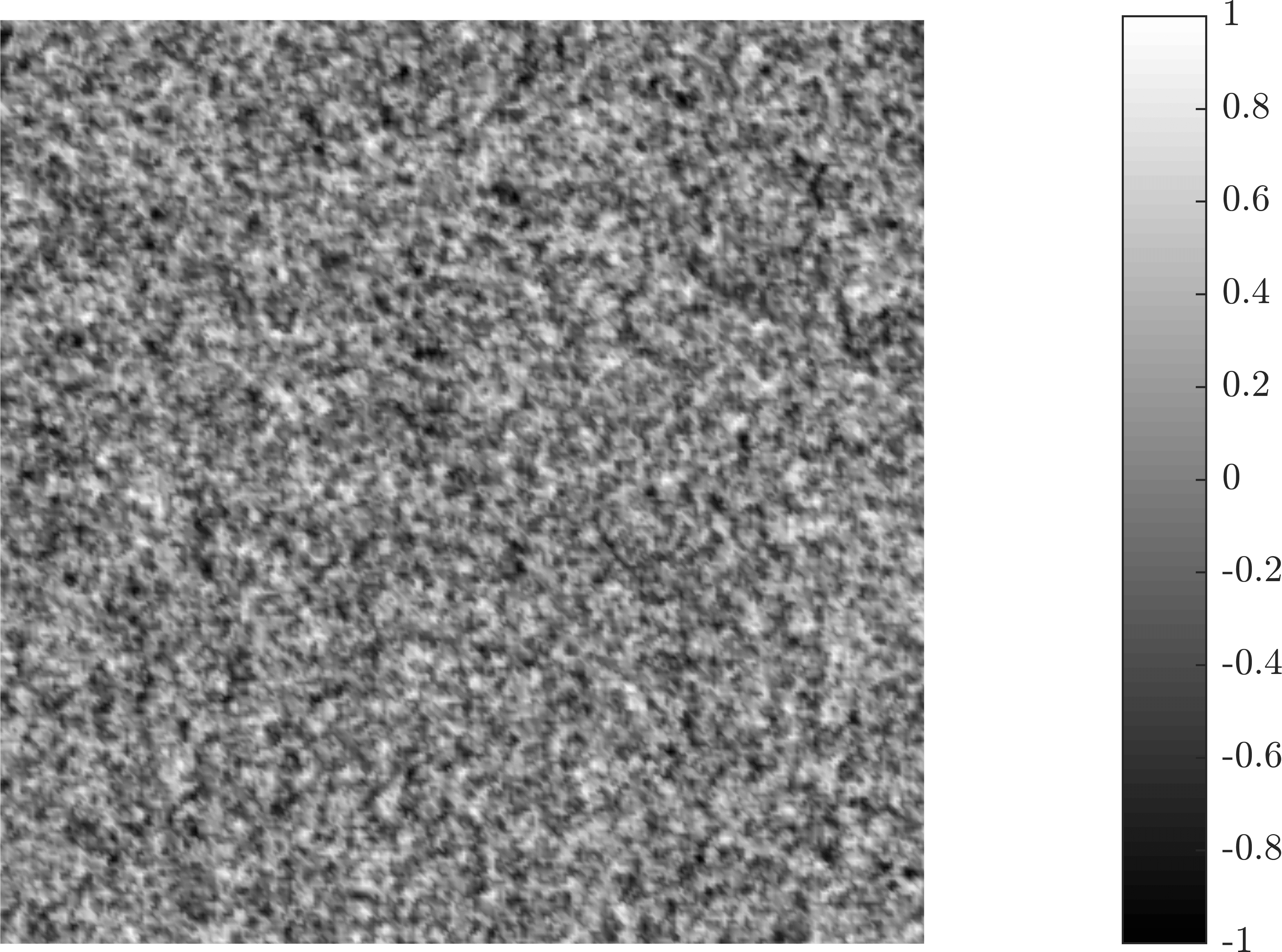}}\hspace{0.5cm}
\subfloat[]{\includegraphics[height=4.8cm]{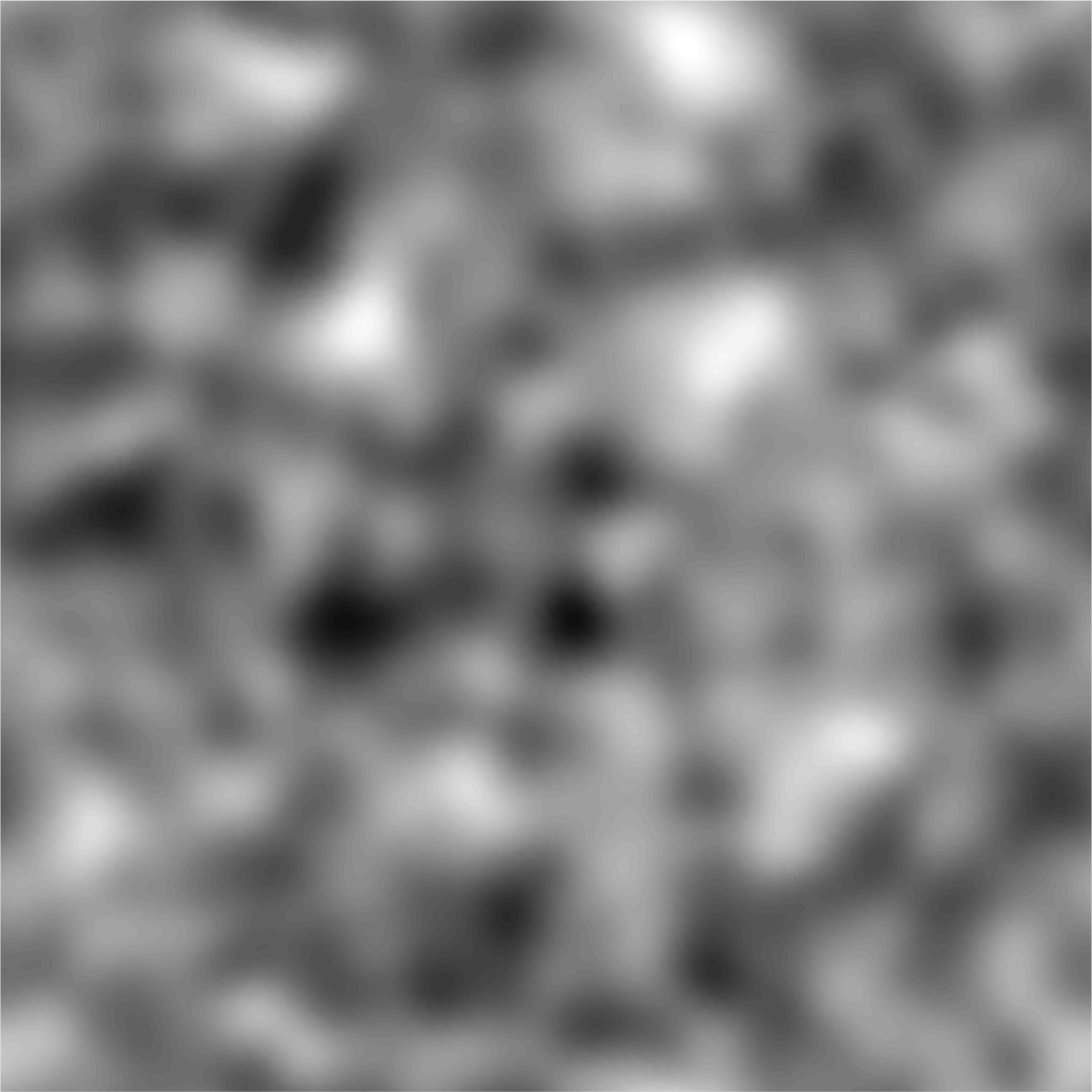}}
\caption{Distribution of the initial plastic distortion ${U}_{13}^{\rm p}/{U}_{13,\rm max}^{\rm p}$. (a) Noisy random microstructure (microstructure A), (b) Smoothed random microstructure (microstructure B).}
\label{fig:Random_micro}
\end{figure}

\aj{\paragraph*{Microstructure A.} First we let the ``noisy'' random microstructure evolve in order to study its equilibrium position. The dislocation density reaches an equilibrium position (see Figure \ref{fig:patterningA} upper left snapshot) consisting in an organized lamellar microstructure mimicking tortuous dislocation cells. The value of the dislocation density $\|{\alpha}\| $ is ten times lower than its initial value, due to an important damping. (The evolution of $\|{\alpha}\|$ is not conservative since the hypotheses of Section \ref{sec:4.1} are not met in the coupled case). The equilibrated microstructure is then subjected to an increasing macroscopic strain ${\bar{\varepsilon}}_{13}$. The results are represented in Figure \ref{fig:patterningA} at several strains: ${\bar{\varepsilon}}_{13} =0$ (equilibrium), ${\bar{\varepsilon}}_{13} =1.3\times 10^{-5}$, ${\bar{\varepsilon}}_{13} =2.2\times 10^{-5}$ and ${\bar{\varepsilon}}_{13} =3 \times 10^{-5}$. The ``lamellar'' microstructure is followed by a ``globular'' microstructure made of dislocation loops separated by thin walls. This type of microstructure evolution, obtained from a random distribution of plastic distortion, resembles the formation of dislocation cells. This apparent dislocation patterning seems very similar to the formation of crystal grains. However, the cells boundaries do not act here as classical grain boundaries where dislocation loops can stack up. Indeed, if the loading is increased, dislocations loops will continue interacting and will annihilate in absence of dislocation nucleation and no pile-up is observed.}
 
\begin{figure}[!ht]
\centering
\includegraphics[width=11cm]{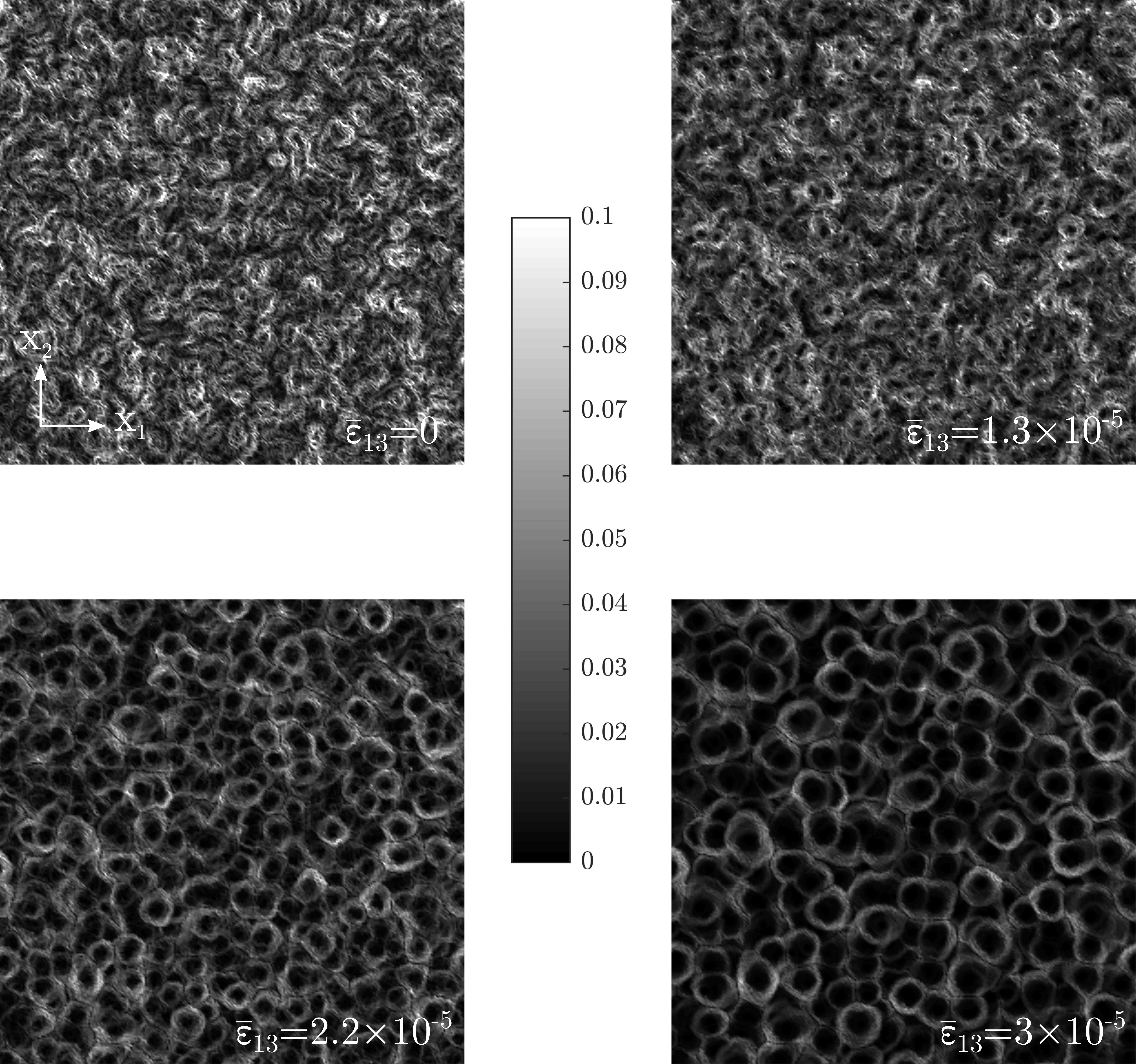}
\caption{\aj{Evolution of the dislocation density $\|{\alpha}\| / \|{\alpha}^0\|_{\rm max}$ of an initially noisy random microstructure (microstructure A) subjected to an increasing macroscopic strain.}}
\label{fig:patterningA}
\end{figure}

\paragraph*{Microstructure B.} \aj{Then we consider the case of the ``smooth microstructure''. Again, the dislocation density reaches an equilibrium position (see Figure \ref{fig:patterningB} upper left snapshot) consisting in an organized lamellar microstructure. The initial cells are more apparent due to a bigger size. It is worth noting that no damping of $\|{\alpha}\| $ is observed, in contrast with the noisy microstructure. The equilibrated microstructure is then subjected to an increasing macroscopic strain ${\bar{\varepsilon}}_{13}$. The results are represented in Figure \ref{fig:patterningB} at several strains: ${\bar{\varepsilon}}_{13} =0$ (equilibrium), ${\bar{\varepsilon}}_{13} =1.3\times 10^{-5}$, ${\bar{\varepsilon}}_{13} =2.2\times 10^{-5}$ and ${\bar{\varepsilon}}_{13} =3 \times 10^{-5}$. Again, the ``lamellar'' microstructure is followed by a ``globular'' microstructure made of dislocation loops separated by thin walls. In that case, the formation of dislocation cells is more patent because they are bigger. This microstructure is very similar to crystal grains in which dislocation loops grow. Again, no pile-up is observed due to dislocation annihilation between loops of neighboring cells.}\\

\begin{figure}[!ht]
\centering
\includegraphics[width=11cm]{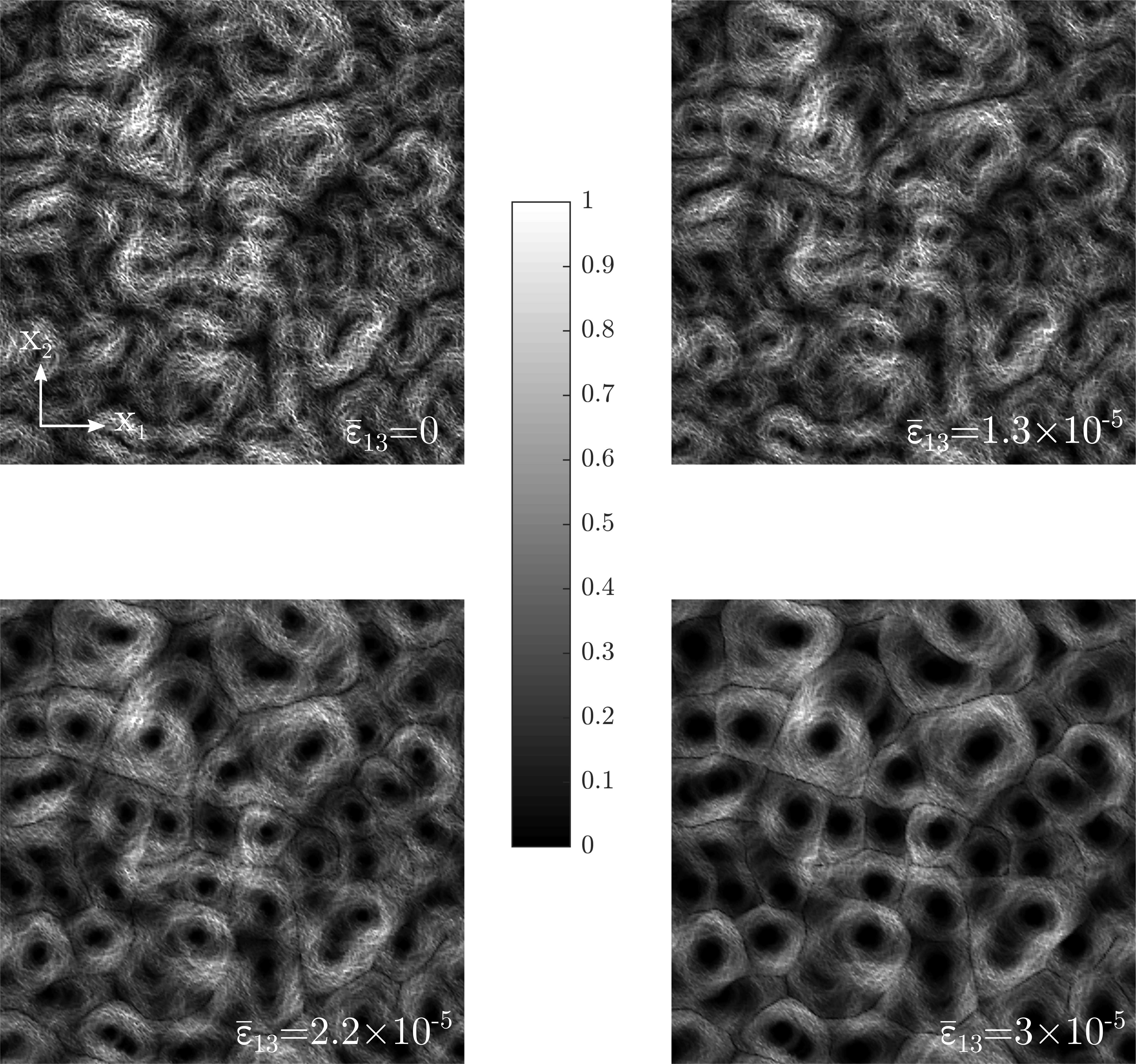}
\caption{\aj{Evolution of the dislocation density $\|{\alpha}\| / \|{\alpha}^0\|_{\rm max}$ of an initially smooth random microstructure (microstructure B) subjected to an increasing macroscopic strain.}}
\label{fig:patterningB}
\end{figure}
 
\aj{In both cases, it is worth noting that strain unloading, up to a nil macroscopic stress, permits to stop the evolution of the microstructure and thus leads to a stabilization of the pattern.}

\section{{Conclusion}}
{The aim of this work was to investigate dislocation-mediated plasticity using the Field Dislocation Mechanics theory. First, \aj{the mesoscale FDM theory was recalled} which has permitted to clearly identify two distinct problems to be solved, the static problem consisting in the determination of the local stress field for a given dislocation density (elliptic equation), and the evolution problem consisting in the transport of the dislocation density (hyperbolic equation). An efficient numerical integration procedure was then proposed. The static problem was solved in a general case using the FFT-based scheme proposed by \cite{brenner_numerical_2014}. The evolution problem, consisting in a vectorial tridimensional Hamilton-Jacobi hyperbolic equation, was solved in a simplified \aj{layer} case using a high resolution Godunov-type scheme. Model problems were finally considered in order to investigate the predictions of the theory. First, uncoupled problems with constant velocity were {explored}: the numerical scheme considered has permitted to reproduce accurately physical phenomena such as the annihilation of dislocations and the expansion of a dislocation loop. Then, \aj{the FDM theory was applied to coupled problems in order to investigate several problems of dislocation-mediated plasticity. In a model problem of interactions between a dislocation and precipitates, the formation of residual dislocation loops around the precipitates has been observed. Finally, the evolution of random microstructures has been studied as a possible way to access dislocation patterning.}\\

{The present work permits to confirm the expectations funded in the Field Dislocation Mechanics theory for predicting several mechanisms of dislocation-mediated plasticity \citep{acharya_new_2010}. The present formulation is not complete since only one component of the plastic distortion tensor was considered. Future developments concerning the numerical integration of the 3-d FDM theory are thus necessary to tackle more general problems of plasticity. A future important task will \aj{notably} consist in developing \aj{numerical algorithms to solve vectorial multi-dimensional Hamilton-Jacobi equations. This would permit to develop a computational implementation of the 3-d FDM theory whose outcome would be a complete description of multiple slips, anisotropy, arbitrary loadings and complex polycrystalline microstructures.}


\aj{\section*{Acknowledgments}
Fruitful discussions with A. Acharya are gratefully acknowledged. The authors would like to thank the two reviewers for their valuable comments and suggestions.}

\appendix

\section{Plastic dissipation and driving force}\label{ap:PKF}

{
The identification of the driving force $\mathbf{F}$ from the expression of the intrinsic dissipation \citep{acharya_driving_2003} is briefly recalled
and its components are given as a function of $\boldsymbol{\sigma}$ and $\mathbf{U}^{\rm p}$ components.\\
By definition (Clausius-Duhem inequality), the intrinsic dissipation is defined by
\begin{equation}\label{eq:dissip}
\mathcal{D}=\int_\Omega (\boldsymbol{\sigma}:{\nabla}\dot{ \tou{u}}-\dot w)
 \dd v
\end{equation}
with $w$ the volumic density of free (stored) energy.
\aj{With the constitutive assumption for the free energy
{$w=\frac{1}{2} \bm{\epsilon}^e:\mathbf{C}:\bm{\epsilon}^e + G(\uu{U}^{\rm p}) $}, the intrinsic dissipation reads
\begin{align}
\mathcal{D}&=\int_\Omega \left(\bm{\sigma}- \frac{\partial G}{\partial \uu{U}^\abr{p}} \right):\dot{\mathbf{U}}^\abr{p} \dd v=
\int_\Omega \left(\bm{\sigma}- \frac{\partial G}{\partial \uu{U}^\abr{p}} \right):\left(\boldsymbol{\alpha}\times\mathbf{V}\right)
\dd v\\[2ex]
&=\int_\Omega
-\left(\bm{\sigma}- \frac{\partial G}{\partial \uu{U}^\abr{p}} \right):\left[(\boldsymbol{\alpha}\otimes\mathbf{V}): \boldsymbol{\epsilon}\right]\dd v
=\int_\Omega -\left[\boldsymbol{\epsilon}:\left(\left(\bm{\sigma}- \frac{\partial G}{\partial \uu{U}^\abr{p}} \right).\boldsymbol{\alpha}
\right) \right].\mathbf{V}\dd v
\end{align}
The driving force associated to the velocity field $\mathbf{V}$ can thus be defined as
\begin{equation}
  \uu{F} = -\boldsymbol{\epsilon}:\left(\left(\bm{\sigma}- \frac{\partial G}{\partial \uu{U}^\abr{p}} \right).\boldsymbol{\alpha}
\right),
\end{equation}
where $\boldsymbol{\epsilon}$ is the Levi-Civita tensor whose components are given by
\begin{equation}
\epsilon_{ijk} =
  \begin{cases}
         +1 & \text{if } (i,j,k) \text{ is } (1,2,3), (2,3,1), \text{ or } (3,1,2), \\
         -1 & \text{if } (i,j,k) \text{ is } (3,2,1), (1,3,2), \text{ or } (2,1,3), \\
    \;\;\,0 & \text{if } i = j, \text{ or } j = k, \text{ or } k = i.
\end{cases}
\end{equation}}

{Besides, a constitutive assumption, ensuring the positivity of the dissipation
$\mathcal{D}$, has to be made for the velocity field $\mathbf{V}$ (see Section \ref{sec:VelocityLaw}).}

\bibliographystyle{model2-names}
\bibliography{ms}

\end{document}